\documentclass{amsart}
\usepackage{latexsym}
\usepackage{amsfonts}
\usepackage{amssymb}
\usepackage{graphicx}
\usepackage{graphicx}

\newtheorem{theorem}{Theorem}[section]
\newtheorem{corollary}[theorem]{Corollary}
\newtheorem{lemma}[theorem]{Lemma}
\newtheorem{proposition}[theorem]{Proposition}
\theoremstyle{definition}

\newtheorem{example}{Example}[section]
\theoremstyle{remark}
\newtheorem{remark}{Remark}[section]
\numberwithin{equation}{section}

\newcommand{\Lg}{\mathfrak{g}}
\newcommand{\Lh}{\mathfrak{h}}
\newcommand{\Lk}{\mathfrak{k}}
\newcommand{\Lp}{\mathfrak{p}}
\newcommand{\La}{\mathfrak{a}}
\newcommand{\Ld}{\mathfrak{d}}
\newcommand{\Lb}{\mathfrak{b}}
\newcommand{\Ll}{\mathfrak{l}}
\newcommand{\Lm}{\mathfrak{m}}
\newcommand{\Ln}{\mathfrak{n}}
\newcommand{\Lt}{\mathfrak{t}}
\newcommand{\Lu}{\mathfrak{u}}
\newcommand{\Lo}{\mathfrak{o}}
\newcommand{\Ly}{\mathfrak{y}}
\newcommand{\gl}{\mathfrak{gl}}
\newcommand{\Lsl}{\mathfrak{sl}}
\newcommand{\Lsp}{\mathfrak{sp}}
\newcommand{\Lsu}{\mathfrak{su}}
\newcommand{\Lso}{\mathfrak{so}}

\newcommand{\ad}{\mathrm{ad}}
\newcommand{\Ad}{\mathrm{Ad}}
\newcommand{\RE}{\mathrm{Re}}
\newcommand{\IM}{\mathrm{Im}}
\newcommand{\A}{\mathrm{A}}
\newcommand{\z}{\mathrm{z}}
\newcommand{\diag}{\mathrm{diag}}
\newcommand{\tr}{\mathrm{tr}}
\newcommand{\alg}{\mathrm{alg}}
\newcommand{\grp}{\mathrm{grp}}
\newcommand{\ii}{\mathbf{i}}
\newcommand{\jj}{\mathbf{j}}
\newcommand{\kk}{\mathbf{k}}
\newcommand{\ee}{\mathbf{e}}
\newcommand{\KK}{\mathbb{K}}
\newcommand{\ZZ}{\mathbb{Z}}
\newcommand{\NN}{\mathbb{N}}
\newcommand{\RR}{\mathbb{R}}
\newcommand{\CC}{\mathbb{C}}
\newcommand{\HH}{\mathbb{H}}
\newcommand{\HC}{\HH_{\CC}}
\newcommand{\D}{\displaystyle}

\begin{document}

\title[]
{A Generalization of random matrix ensemble II: concrete examples
and integration formulae$^1$}
\author[]{Jinpeng An}
\address{School of mathematical science, Peking University, Beijing, 100871, P. R. China}
\email{anjinpeng@math.pku.edu.cn}

\author[]{Zhengdong Wang}
\address{School of mathematical science, Peking University, Beijing, 100871, P. R. China }
\email{zdwang@pku.edu.cn}

\author[]{Kuihua Yan}
\address{School of mathematics and physics, Zhejiang Normal
University, Zhejiang Jinhua, 321004, P. R. China }
\email{yankh@zjnu.cn}

\begin{abstract}
According to the classification scheme of the generalized random
matrix ensembles, we present various kinds of concrete examples of
the generalized ensemble, and derive their joint density functions
in an unified way by one simple formula which was proved in
\cite{AWY}. Particular cases of these examples include Gaussian
ensemble, chiral ensemble, new transfer matrix ensembles, circular
ensemble, Jacobi ensembles, and so on. The associated integration
formulae are also given, which are just many classical integration
formulae or their variation forms.
\end{abstract}
\maketitle


\footnotetext[1]{This work is supported by the 973 Project
Foundation of China ($\sharp$TG1999075102).
}

\vskip 0.5cm
\section{Introduction}

\vskip 0.5cm Guided by Dyson's idea in \cite{Dy}, many authors
have investigated the method of deriving the joint density
functions of various kinds of random matrix ensembles in terms of
Riemannian symmetric spaces. One of the most important works in
this direction was made by Due\~{n}ez \cite{Du}, in which the
joint density functions for the circular ensemble and various
kinds of Jacobi ensembles was obtained using an integration
formula associated with the $KAK$ decomposition of compact Lie
group, according to Cartan's classification of compact irreducible
Riemannian symmetric spaces. The achievement of this direction was
summarized by the excellent review article of Caselle and Magnea
\cite{CM}.

\vskip 0.3cm This is a sequel paper of \cite{AWY}, in which a
generalization of the random matrix ensemble was defined. First we
give a sketch of the content of \cite{AWY}. Suppose a Lie group
$G$ acts on an $n$-dimensional Riemannian manifold $X$ by $\sigma:
G\times X\rightarrow X$, and suppose the induced Riemannian
measure $dx$ is $G$-invariant. Let $Y$ be a closed submanifold of
$X$ with the induced Riemannian measure $dy$, and let $K=\{g\in
G:\sigma_g(y)=y, \forall y\in Y\}$. Define the map $\varphi:
G/K\times Y\rightarrow X$ by $\varphi([g],y)=\sigma_g(y)$. Let
$X_\z \subset X$, $Y_\z \subset Y$ be closed subsets of measure
zero in $X$ and $Y$, respectively. Denote $X' = X \setminus X_\z$,
$Y' = Y \setminus Y_\z$. Suppose the following conditions hold.

\vskip 0.3cm {\flushleft{\bf(a)}} \quad (\emph{invariance
condition}) \quad $X' = \D\bigcup_{y\in Y'} O_y$.

\vskip 0.1cm {\flushleft{\bf(b)}} \quad (\emph{transversality
condition}) \quad $T_y X=T_y O_y\oplus T_y Y$, \quad $\forall y\in
Y'$.

\vskip 0.3cm {\flushleft{\bf(c)}} \quad (\emph{dimension
condition}) \quad $\mathrm{dim}G_y=\mathrm{dim}K, \quad \forall
y\in Y'$.

\vskip 0.3cm {\flushleft{\bf(d)}} \quad (\emph{orthogonality
condition}) \quad $T_yY\perp T_yO_y$, \quad $\forall y\in Y'$.

\vskip 0.3cm {\flushleft Suppose} $d\mu$ is a $G$-invariant smooth
measure on $G/K$, and suppose $p(x)$ is a $G$-invariant smooth
function on $X$. Then the system $(G,\sigma,X,p(x)dx,Y,dy)$ is
called a generalized random matrix ensemble. $X$ and $Y$ are
called the integration manifold and the eigenvalue manifold,
respectively. It is proved that there is a quasi-smooth measure
$d\nu$ on $Y$, which is called the generalized eigenvalue
distribution, such that $\varphi^*(p(x)dx)=d\mu d\nu$. We write
$d\nu$ as the form $d\nu(y)=\mathcal{P}(y)dy=p(y)J(y)dy$, where
$\mathcal{P}(y)=p(y)J(y)$ is called the generalized joint density
function. For $y\in Y'$ define the map $\Psi_y : \Ll \rightarrow
T_yO_y$ by $\Psi_y(\xi)=\frac{d}{dt}\big|_{t=0}\sigma_{\exp
t\xi}(y)$, $\forall\xi\in\Ll$, where $\Ll$ is a linear subspace of
$\Lg$ such that $\Lg=\Lk\oplus\Ll$. The main result in \cite{AWY}
is
\begin{equation}\label{E:main}
J(y)=C|\det\Psi_y|,\quad C=|\det((d\pi)_e|_\Ll)|^{-1}.
\end{equation}
If in addition the following covering condition holds:

\vskip 0.3cm {\flushleft{\bf(e)}} \quad (\emph{covering
condition}) \quad The map $\varphi: G/K\times Y'\rightarrow X'$ is
a $d$-sheeted covering map, with $d<+\infty$.

\vskip 0.3cm {\flushleft Then} it is also proved in \cite{AWY}
that {\small\begin{equation}\label{E:main2} \int_X f(x)p(x)
dx=\frac{1}{d}\int_Y\left(\int_{G/K}f(\sigma_g(y))d\mu([g])\right)
d\nu(y)
\end{equation}}
for all $f\in C^\infty(X)$ with $f\geq0$ or with $f\in
L^1(X,p(x)dx)$. \cite{AWY} also give a classification scheme of
the generalized ensembles, that is, the linear ensemble, the
nonlinear noncompact ensemble, the compact ensemble, the group
ensemble, the algebra ensemble, the pseudo-group ensemble, and the
pseudo-algebra ensemble.

\vskip 0.3cm We should point out that though the proof of Formula
\eqref{E:main} is not difficult, it provides a direct and unified
way to compute the joint density functions for various kinds of
random matrix ensembles. In this paper we will show that all the
classical ensembles are particular cases of the generalized
ensemble, and the corresponding density functions can be derived
directly from \eqref{E:main}. The density functions for some new
examples of the generalized ensemble can also derived form
\eqref{E:main} explicitly.

\vskip 0.3cm According to the classification scheme of the
generalized ensembles which was given in \cite{AWY}, we will
present various kinds of concrete examples of generalized
ensemble, and derive their joint density functions explicitly. The
associated integration formulae will also be given. In \S 2 we
will consider the linear ensemble, and present examples associated
with $GL(n,\KK), O(m,n)_0, U(m,n)$, and $Sp(m,n)$. Particular
cases associated with $GL(n,\KK)$ is the Gaussian ensemble, and
particular cases associated with $O(m,n)_0, U(m,n)$, and $Sp(m,n)$
are the chiral ensembles (which are also called Laguerre
ensembles). The four classes of the BdG ensemble and the two
classes of the $p$-wave ensemble are also particular cases of the
linear ensemble. The associated integration formula is just the
integration formula for the Cartan decomposition of reductive Lie
algebra in \cite{He}. In \S 3 examples associated with $GL(n,\KK),
O(m,n)_0, U(m,n)$, and $Sp(m,n)$ of the nonlinear noncompact
ensemble will be presented. Particular cases associated with
$GL(n,\KK)$ are the so-called new transfer matrix ensembles. The
three cases of the transfer matrix ensemble are also particular
cases of the nonlinear noncompact ensemble. The associated
integration formula is a variation form of the integration formula
for Riemannian symmetric space of noncompact type in \cite{He}. In
\S 4 we present examples of the compact ensemble associated with
$G_*=GL(n,\KK)$ and $G=SO(m+n), U(m+n), Sp(m+n)$. Particular case
associated with $GL(n,\KK)$ is the circular ensemble, and
particular cases associated with $SO(m+n), U(m+n)$, and $Sp(m+n)$
are Jacobi ensembles. The associated integration formula is a
variation form of the integration formula for Riemannian symmetric
space of compact type in \cite{He}. \S 5 and \S 6 will be devoted
to various examples of the group ensemble and the algebra
ensemble. Examples associated with $U(n), SO(2n+1), Sp(n), SO(2n),
SL(n,\CC), Sp(n,\CC), SO(2n,\CC)$, and $SO(2n+1,\CC)$ will be
presented. We note that in some literatures, the group $Sp(n)$ is
denoted by $USp(2n)$. Here we follow the notation in Knapp
\cite{Kn}. The associated integration formulae will be the Weyl
integration formula for compact groups and the Harish-Chandra's
integration formula for complex semisimple Lie groups, as well as
their Lie algebra version. In \S 7 we will consider the
pseudo-group ensemble and the pseudo-algebra ensemble, presenting
examples associated with $SL(2,\RR)$ and $GL(n,\RR)$. As a
corollary of the associated integration formula, we will recover
Harish-Chandra's integration formula for real reductive group and
its Lie algebra version.


\vskip 0.5cm
\section{Linear ensembles}

\vskip 0.5cm In this section we consider the linear ensemble. Let
$G$ be a real reductive Lie group with Lie algebra $\Lg$. Then $G$
admits a global Cartan involution $\Theta$, which induces a Cartan
involution $\theta$ of $\Lg$ with the associated Cartan
decomposition $\Lg=\Lk\oplus\Lp$. Let $K=\{g\in G:\Theta(g)=g\}$,
which is a maximal compact subgroup of $G$ with Lie algebra $\Lk$.
Let $\La$ be a maximal abelian subspace of $\Lp$, and let
$A=\exp(\La)$ be the connected subgroup of $G$ with Lie algebra
$\La$. Then $\Lp=\bigcup_{k\in K}\A_k(\La), P=\bigcup_{k\in
K}\sigma_k(A)$. Let $ M=\{k\in K:\A_k(\eta)=\eta, \forall
\eta\in\La\}=\{k\in K:\sigma_k(a)=a, \forall a\in A\}$,
$\Lm=\{\xi\in \Lk:[\xi,\eta]=0,\forall \eta\in\La\}$, then $M$ is
a closed subgroup of $K$ with Lie algebra $\Lm$. Let $\Sigma$ be
the restricted root system associated with $\La$ with the Weyl
group $W=W(\Sigma)$. For $\lambda\in\Sigma$, let $\Lg_\lambda$ be
the corresponding root space. We choose a notion of positivity in
$\Sigma$ and denote by $\Sigma^+$ the set of positive restricted
roots. There is a nondegenerate symmetric bilinear form $B$ on
$\Lg$ which is invariant under $\theta$ and $\Ad(g)$ for all $g\in
G$, and satisfies that $\Lk$ and $\Lp$ are orthogonal under $B$,
$B|_\Lk$ is negative definite, and $B|_\Lp$ is positive definite.
So $\langle\xi,\eta\rangle=-B(\xi,\theta\eta)$ defines an inner
product on $\Lg$. We write $\Lb=\La^\perp$ in $\Lp$ and
$\Ll=\Lm^\perp$ in $\Lk$, then
$\Lb\oplus\Ll=\bigoplus_{\lambda\in\Sigma}\Lg_\lambda$. For each
$\lambda\in\Sigma^+$ we choose an orthogonal basis
$\{\gamma_{\lambda,1},\cdots,\gamma_{\lambda,\beta_\lambda}\}$ of
$\Lg_\lambda$ such that $|\gamma_{\lambda,j}|=\frac{\sqrt{2}}{2}$,
where $\beta_\lambda=\dim\Lg_\lambda$. For each
$\gamma_{\lambda,j}$, denotes
$\xi_{\lambda,j}=\gamma_{\lambda,j}+\theta \gamma_{\lambda,j}$,
$\zeta_{\lambda,j}=\gamma_{\lambda,j}-\theta \gamma_{\lambda,j}$,
then $|\xi_{\lambda,j}|=|\zeta_{\lambda,j}|=1$, and we have $
\theta\xi_{\lambda,j}=\xi_{\lambda,j},
\theta\zeta_{\lambda,j}=-\zeta_{\lambda,j}$. So $
\xi_{\lambda,j}\in\Lk\cap\bigoplus_{\lambda\in\Sigma}\Lg_{\lambda}=\Ll,\,
\zeta_{\lambda,j}\in\Lp\cap\bigoplus_{\lambda\in\Sigma}\Lg_{\lambda}=\Lb.
$ But the set
$\{\xi_{\lambda,j},\zeta_{\lambda,j}:\lambda\in\Sigma^+,
j=1,\cdots,\beta_\lambda\}$ is linearly independent, so
$$\{\xi_{\lambda,j}:\lambda\in\Sigma^+,
 j=1,\cdots,\beta_\lambda\}\subset\Ll$$ is an orthonormal basis
for $\Ll$, and $$\{\zeta_{\lambda,j}:\lambda\in\Sigma^+,
j=1,\cdots,\beta_\lambda\}\subset\Lb$$ is an orthonormal basis for
$\Lb$. And then we have
$\dim\Ll=\dim\Lb=\sum_{\lambda\in\Sigma^+}\beta_\lambda$. Let
$P=\exp(\Lp)$, which is a closed submanifold of $G$ satisfies
$T_eP=\Lp$. In fact, $P$ is the identity component of the set
$\{g\in G:\Theta(g)=g^{-1}\}$ (see \cite{AW}). The exponential map
$\exp:\Lp\rightarrow P$ is a diffeomorphism, so we can define its
inverse map $\log:P\rightarrow\Lp$. We also have the global Cartan
decomposition $G=K\times P$. It is known that $\Lp$ is an
invariant subspace of the adjoint action $\Ad|_K$, and $P$ is also
invariant under the conjugate action of $K$. We denote
$\A_k=\Ad(k)|_{\Lp}$ and $\sigma_k(p)=kpk^{-1}$ for $k\in K$ and
$p\in P$.

\vskip 0.3cm In this section we consider the action $\A_k$ of $K$
on $\Lp$. The inner product $\langle\cdot,\cdot\rangle$ induces a
linear Riemannian structure on $\Lp$, which is $K$-invariant under
the action $\A_k$. So it induces a $K$-invariant Riemannian
measure $dX$, which is just the Lebesgue measure on $\Lp$. Let
$p(\xi)$ be a $K$-invariant positive smooth function on $\Lp$,
then $p(\xi)dX(\xi)$ is a $K$-invariant smooth measure. The
Riemannian structure on $\Lp$ also induces a Riemannian measure
$dY$ on $\La$, which is also the Lebesgue measure. Define the map
$\varphi:K/M\times\La\rightarrow\Lp$ by
$\varphi([k],\eta)=\A_k(\eta)$. It is easy to prove that
$\Ad_M(\Ll)\subset\Ll$, so under the natural identification
$(d\pi)_e|_\Ll:\Ll\rightarrow T_{[e]}(K/M)$, the $\Ad_M$-invariant
inner product $\langle\cdot,\cdot\rangle|_\Ll$ on $\Ll$ induces a
$K$-invariant Riemannian structure on $K/M$, and then induces a
$K$-invariant smooth measure $d\mu$ on $K/M$. These are sufficient
for us to form a concrete example of the generalized random matrix
ensemble with integration manifold $\Lp$ and eigenvalue manifold
$\La$.

\begin{theorem}\label{T:linear}
The system $(K,\A,\Lp,p(\xi)dX(\xi),\La,dY)$ is a generalized
random matrix ensemble. Its generalized joint density function
$\mathcal{P}(\eta)=p(\eta)J(\eta)$ is given by
{\small\begin{equation}\label{E:linear}
J(\eta)=\prod_{\lambda\in\Sigma^+}|\lambda(\eta)|^{\beta_{\lambda}}.
\end{equation}}
\end{theorem}

\begin{proof}
For $\eta\in\La$, we consider the map $\Psi_\eta:\Ll\rightarrow
T_\eta O_\eta$ defined by
$\Psi_\eta(\xi)=\frac{d}{dt}\big|_{t=0}A_{\exp t\xi}(\eta)$. We
have
\begin{align*}
\Psi_\eta(\xi_{\lambda,j})
=&\frac{d}{dt}\Big|_{t=0}\A_{\exp t\xi_{\lambda,j}}(\eta)\\
=&[\xi_{\lambda,j},\eta]\\
=&-\lambda(\eta)(\zeta_{\lambda,j}).
\end{align*}
So for
$\eta\in\La\setminus\left(\bigcup_{\lambda\in\Sigma^+}\ker\lambda\right)$,
$\Psi_\eta$ is an isomorphism, hence $T_\eta
O_\eta=Im(\Psi_\eta)=\Lb$. Let
$\Lp_\z=\varphi(K/M,\bigcup_{\lambda\in\Sigma^+}\ker\lambda)$,
then $\Lp_\z$ and
$\La_\z=\Lp_\z\cap\La=\bigcup_{\lambda\in\Sigma^+}\ker\lambda$ are
lower-dimensional sets in $\Lp$ and $\La$, respectively (in the
sense of \cite{Kn}, Section 8.1), thus they have measures zero in
the corresponding spaces.

\vskip 0.3cm Now we check the conditions (a), (b), (c), and (d).
Let $\Lp'=\Lp\setminus \Lp_\z$ and $\La'=\La\cap
\Lp'=\La\setminus\left(\bigcup_{\lambda\in\Sigma^+}\ker\lambda\right)$.
We have shown the condition (a) holds. For $\eta\in \La'$, by the
definition of the Riemannian structure on $\Lp$,
$T_\eta\Lp=\Lp=\La\oplus\Lb=T_\eta\La\oplus T_\eta O_\eta$
orthogonally, so the conditions (b) and (d) hold. For
$\eta\in\La'$, suppose the isotropic subgroup associated with
$\eta$ is $K_\eta$, then
\begin{align*}
\dim K_\eta
=&\dim K-\dim O_\eta\\
=&\dim K-\dim\Lb\\
=&\dim K-\dim\Ll\\
=&\dim M.
\end{align*}
So the condition (c) also holds. This proves the system
$(K,\A,\Lp,p(\xi)dX(\xi),\La,dY)$ is a generalized random matrix
ensemble.

\vskip 0.3cm We have seen above that
$\Psi_\eta(\xi_{\lambda,j})=-\lambda(\eta)(\zeta_{\lambda,j})$ for
each $\lambda\in \Sigma^+$ and $j=1,\cdots,\beta_\lambda$. By
Formula \eqref{E:main}, {\small$$
J(\eta)=C\prod_{\lambda\in\Sigma^+}|\lambda(\eta)|^{\beta_{\lambda}},
$$} where $C=|\det((d\pi)_e|_\Ll)|^{-1}$. But $(d\pi)_e|_\Ll$ is isometric, so $C=1$.
This complete the proof of the theorem.
\end{proof}

From Theorem \ref{T:linear} we know that the generalized
eigenvalue distribution $d\nu$ is given by {\small\begin{equation}
d\nu(\eta)=p(\eta)\prod_{\lambda\in\Sigma^+}|\lambda(\eta)|^{\beta_{\lambda}}dY(\eta).
\end{equation}}
The generalized random matrix ensemble in Theorem \ref{T:linear}
is called \emph{linear ensemble}.

\begin{corollary}\label{C:linearInt}
Let $f\in C^\infty(\Lp)$ satisfies $f\geq0$ or $f\in
L^1(\Lp,p(\xi)dX(\xi))$. Then we have the following integration
formula {\small\begin{equation}\label{E:linearInt} \int_\Lp
f(\xi)p(\xi)dX(\xi)=\frac{1}{|W|}\int_\La\left(\int_{K/M}f(\A_k(\eta))d\mu([k])\right)
d\nu(\eta),
\end{equation}}
where $|W|$ is order of the Weyl group $W$.
\end{corollary}

\begin{proof}
By Formula \eqref{E:main2}, it is sufficient to show that the
covering condition (e) holds, and the covering sheet is $|W|$. For
each $\eta\in\La'$, suppose $\A_k(\eta)=\Ad(k)\eta=\eta$ for some
$k\in K$. Since $\Ad(k)$ is an automorphism of $\Lg$, $\Ad(k)$
must fix $Z_\Lg(\eta)=\Lg_0$. but $\Lp$ is also fixed by $\Ad(k)$,
so $\Ad(k)$ fix $\Lg_0\cap\Lp=\La$, that is $k$ is in the
normalizer $N_K(\La)=\{k\in K:\Ad(k)(\La)=\La\}$ of $\La$. But it
is known that the analytic Weyl group $W(G,A)=N_K(\La)/M$ is
coincides with $W(\Sigma)$ (see \cite{Kn}, Proposition 7.32), so
the action $\Ad(k)$ on $\La$ coincides with some $w\in W$. But
$\eta$ is a regular element and $w(\eta)=\eta$, this force that
$w=1$, that is $k\in Z_K(\La)=M$. This proves that for each
$\eta\in Y'$, the isotropic subgroup $K_\eta=M$. Next, also by the
relation $W(G,A)=W(\Sigma)$, it follows that for each
$\eta\in\La'$, $O_\eta\cap Y'$ has $|W|$ points. By Corollary 3.6
in \cite{AWY}, $\varphi:K/M\times Y'\rightarrow X'$ is a
$|W|$-sheeted covering map. This proves the corollary.
\end{proof}

\begin{remark}
Formula \eqref{E:linearInt} has appeared in Helgason \cite{He}
(Chapter 1, Theorem 5.17). Here we recover it from the viewpoint
of generalized random matrices.
\end{remark}

\begin{example}\label{Ex:GL(n,K)linear}
Let $G=GL(n,\KK)$, where $\KK$ is $\RR$, $\CC$, or $\HH$. Then $G$
is real reductive, when we view $GL(n,\CC)$ and $GL(n,\HH)$ as
real Lie groups. The Cartan involution of the Lie algebra
$\Lg=\gl(n,\KK)$ can be chosen as $\theta(\xi)=-\xi^*$, where the
symbol ``$\xi^*$" means the transpose of $\xi$ when $\KK=\RR$, and
the conjugate transpose when $\KK=\CC$ or $\HH$. The corresponding
Cartan decomposition is
$\Lk=\{\xi\in\Lg\Ll(n,\mathbb{K}):\xi^*=-\xi\}$,
$\Lp=\{\xi\in\Lg\Ll(n,\mathbb{K}):\xi^*=\xi\}$. The space
$\La=\{\eta=\diag(x_1,\cdots,x_n):x_k\in\mathbb{R}\}$ is a maximal
abelian subspace of $\Lp$ in each of the three cases. The
corresponding global Cartan involution of $GL(n,\KK)$ is
$\Theta(g)=(g^*)^{-1}$, and the maximal compact subgroup $K=\{g\in
G:\Theta(g)=g\}$ is $O(n)$, $U(n)$, or $Sp(n)$ when $\KK$ is
$\RR$, $\CC$, or $\HH$, respectively. Let $\ee_r\in\La^*$ denotes
$\ee_r(\diag(x_1,\cdots,x_n))=x_r$ for each $1\leq r\leq n$, then
one can choose the positive restricted root system as
$\Sigma^+=\{\ee_r-\ee_s:1\leq r<s\leq n\},$ and
$\beta_{\ee_r-\ee_s}=\dim\Lg_{\ee_r-\ee_s}=\beta$, where
$\beta=1,2,$ or $4$ when $\mathbb{K}$ is $\mathbb{R}$,
$\mathbb{C}$, or $\mathbb{H}$, respectively. Let $p(\xi)$ be a
$K$-invariant positive smooth function on $\Lp$. Then by Theorem
\ref{T:linear}, the density function
$\mathcal{P}(\eta)=p(\eta)J(\eta)$ for the linear ensemble
$(K,\A,\Lp,p(\xi)dX(\xi),\La,dY)$ is determined by
{\small\begin{equation} J(\eta)=\prod_{1\leq r<s\leq
n}|x_r-x_s|^\beta.
\end{equation}}
If the function $p(\xi)$ is of the particular form
$p(\xi)=\exp(-a\tr\xi^2+b\tr\xi+c)$ such that $p(\xi)dX(\xi)$ is a
probability measure, the linear ensemble
$(K,\A,\Lp,p(\xi)dX(\xi),\La,dY)$ is just the Gaussian orthogonal,
unitary, and symplectic ensembles. Thus we recover the joint
density functions for the three cases of Gaussian ensemble from
the viewpoint of generalized random matrix ensemble.\qed
\end{example}

\begin{example}\label{Ex:SO(m,n)linear}
Let $G=O(m,n)_0$, $U(m,n)$, or $Sp(m,n)$, which are all real
reductive. These groups are defined to be the connected component
of
$$\{g\in GL(m+n,\KK):g^*I_{m,n}g=I_{m,n}\},$$ where $\KK=\RR, \CC$
or $\HH$, respectively, where $I_{m,n}={\SMALL
\begin{pmatrix}
I_m & 0\\
0 & -I_n
\end{pmatrix}}.$ Without loss of
generality, we may assume $m\geq n$. The Lie algebras $\Lo(m,n),
\Lu(m,n)$, and $\Lsp(m,n)$ of the three groups are
\begin{align*}
\Lg=&\{\xi\in\gl(m+n,\KK):\xi^*I_{m,n}+I_{m,n}\xi=0\}\\
=&\left\{{\SMALL
\begin{pmatrix}
A & B\\
B^* & D
\end{pmatrix}}
:A+A^*=0,D+D^*=0\right\}.
\end{align*}
The Cartan involution of $\Lg$ can be chosen as
$\theta(\xi)=-\xi^*$, and the corresponding Cartan decompositions
is $ \Lk=\left\{{\SMALL
\begin{pmatrix}
A & 0\\
0 & D
\end{pmatrix}}\right\},
\Lp=\left\{{\SMALL
\begin{pmatrix}
0 & B\\
B^* & 0
\end{pmatrix}}\right\}.
$ Let $E_{rs}$ denotes the $(m+n)$-by-$(m+n)$ matrix with $1$ at
the $(r,s)$-th entry and $0$ elsewhere. Then one can easily
checked the space {\small\begin{equation}\label{E:a}
\La=\left\{\eta=
\sum_{k=1}^nx_k(E_{m-k+1,m+k}+E_{m+k,m-k+1}):x_k\in\RR\right\}
\end{equation}}
is a maximal abelian subspace of $\Lp$ in each of the three cases.
The corresponding global Cartan involution is
$\Theta(g)=(g^*)^{-1}$, and the maximal compact subgroup $K=\{g\in
G:\Theta(g)=g\}=S(O(m)\times O(n))$ (which means the subgroup of
$O(m)\times O(n)$ consists of elements with determinant $1$),
$U(m)\times U(n)$, or $Sp(m)\times Sp(n)$ when $G=O(m,n)_0$,
$U(m,n)$, or $Sp(m,n)$, respectively. Let $\ee_r\in\La^*$ denotes
$\ee_r(\eta)=x_r$ for each $1\leq r\leq n$, then one can choose
the positive restricted root system as
\begin{equation}\label{E:root:O(m,n)}
\Sigma^+= \{\ee_r\pm\ee_s:1\leq r<s\leq n\}
\cup\{\ee_r,2\ee_r:1\leq r\leq n\},
\end{equation}
and it can be shown that $\beta_{\ee_r\pm\ee_s}=\beta$,
$\beta_{\ee_r}=\beta(m-n)$, and $\beta_{\ee_r}=\beta-1$, where
$\beta=1,2$, or $4$ when $G=O(m,n)_0$, $U(m,n)$, or $Sp(m,n)$ (if
$\beta_\lambda=0$ for some $\lambda\in\Sigma^+$, the root
$\lambda$ should be omitted). By Theorem \ref{T:linear}, we can
compute the factor $J(\eta)$ as
{\small\begin{equation}\label{E:(m,n)linear}
J(\eta)=2^{(\beta-1)n}\prod_{1\leq r<s\leq
n}|x_r^2-x_s^2|^\beta\prod_{r=1}^n|x_r|^{\beta(m-n+1)-1}.
\end{equation}}
Let $p(\xi)$ be a $K$-invariant positive smooth function on $\Lp$,
then the density function $\mathcal{P}(\eta)=p(\eta)J(\eta)$ for
the three cases of the linear ensemble
$(K,\A,\Lp,p(\xi)dX(\xi),\La,dY)$ is determined by Formula
\eqref{E:(m,n)linear}. If the function $p(\xi)$ is chosen of the
particular form $p(\xi)=\exp(-a\tr(\xi\xi^*))$ such that
$p(\xi)dX(\xi)$ is a probability measure, the linear ensemble
$(K,\A,\Lp,p(\xi)dX(\xi),\La,dY)$ is just the chiral orthogonal,
unitary, and symplectic ensembles (see \cite{CM}). \qed
\end{example}

\vskip 0.3cm The four classes of the BdG ensemble and the two
classes of the $p$-wave ensemble are also particular cases of the
linear ensemble. For the lack of space, we only point out what the
corresponding groups $G$ and $K$ are. The reader can easily obtain
the other objects and derive their joint density functions from
Theorem \ref{T:linear}. For the BdG ensembles, $G=SO(4n,\CC),
Sp(n,\CC), SO^*(4n)$, and $Sp(n,\RR)$, the corresponding
$K=SO(4n), Sp(n), U(2n)$, and $U(n)$, respectively. For the
$p$-wave ensembles, $G=SO(2n+1,\CC)$ and $SO^*(4n+2)$, the the
corresponding $K=SO(2n+1)$ and $U(2n+1)$.

\vskip 0.5cm
\section{Nonlinear noncompact ensembles}

\vskip 0.5cm In some sense, the nonlinear noncompact ensemble is
the nonlinear version of the linear ensemble. But something will
be different. Let $G$ be a reductive Lie group, and keep the
notations at the beginning of \S 2. Recall that the group $K$ acts
on $P$ by $\sigma_k(p)=kpk^{-1}$. The inner product
$\langle\cdot,\cdot\rangle$ induces a $G$-left invariant and
$K$-bi-invariant Riemannian structure on $G$, and then induces
Riemannian structures on $P$ and $A$ as well as the Riemannian
measures $dx$ and $da$ on $P$ and $A$, respectively. Since the
induced Riemannian structure on $P$ is $K$-invariant, the measure
$dx$ on $P$ is also $K$-invariant. As in the previous section, the
inner product $-B|_\Ll$ induces a $K$-invariant Riemannian
structure on $K/M$, then induces a $K$-invariant Riemannian
measure $d\mu$. Let $p(x)$ be a $K$-invariant positive smooth
function on $P$. Define the map $\varphi:K/M\times A\rightarrow P$
by $\varphi([k],a)=\sigma_k(a)$. Then we can construct the
\emph{nonlinear noncompact ensemble} with integration manifold $P$
and eigenvalue manifold $A$ as follows.

\begin{theorem}\label{T:nonlinear}
Let the objects be as above. Then the system
$(K,\sigma,P,p(x)dx,A,da)$ is a generalized random matrix
ensemble. Its generalized joint density function
$\mathcal{P}(a)=p(a)J(a)$ is given by
{\small\begin{equation}\label{E:nonlinear}
J(a)=2^{\dim\Ll}\prod_{\lambda\in\Sigma^+}\left(\Bigl|
\sinh\frac{\lambda(\eta)}{2}\Bigr|
\sqrt{\cosh{\lambda(\eta)}}\right)^{\beta_{\lambda}},
\end{equation}}
where $\eta=\log a$.
\end{theorem}

\begin{proof}
For $a\in A$, consider the map $\Psi_a:\Ll\rightarrow T_aO_a,
\Psi_a(\xi)=\frac{d}{dt}\big|_{t=0}\sigma_{\exp t\xi}(a)$. Then we
have
{\small\begin{align*}
\Psi_a(\xi_{\lambda,j})
=&\frac{d}{dt}\Big|_{t=0}e^{t\xi_{\lambda,j}}ae^{-t\xi_{\lambda,j}}\\
=&(dl_a)\frac{d}{dt}\Big|_{t=0}e^{t\Ad(a^{-1})\xi_{\lambda,j}}e^{-t\xi_{\lambda,j}}\\
=&(dl_a)\left(\Ad(a^{-1})\xi_{\lambda,j}-\xi_{\lambda,j}\right)\\
=&(dl_a)\left(e^{-\ad\eta}\xi_{\lambda,j}-\xi_{\lambda,j}\right)\\
=&(dl_a)\left(e^{-\lambda(\eta)}\gamma_{\lambda,j}+e^{\lambda(\eta)}\theta\gamma_{\lambda,j}-\xi_{\lambda,j}\right)\\
=&(dl_a)\big((-\sinh\lambda(\eta))\zeta_{\lambda,j}+
(\cosh\lambda(\eta)-1)\xi_{\lambda,j}\big).
\end{align*}}
Since $dl_a$ is isometric, {\small\begin{align*}
|\Psi_a(\xi_{\lambda,j})|
=&|(-\sinh\lambda(\eta))\zeta_{\lambda,j}+
(\cosh\lambda(\eta)-1)\xi_{\lambda,j}|\\
=&\sqrt{\sinh^2\lambda(\eta)+(\cosh\lambda(\eta)-1)^2}\\
=&2\left|\sinh\frac{\lambda(\eta)}{2}\,\right|
\sqrt{\cosh{\lambda(\eta)}}.
\end{align*}}
So if $\lambda(\eta)\neq0$, that is $a\notin\exp(\ker\lambda)$,
then $|\Psi_a(\xi_{\lambda,j})|\neq0$. Let
$A_\z=\bigcup_{\lambda\in\Sigma^+}\exp(\ker\lambda)$, then for
$a\in A'=A\backslash A_\z$, $\Psi_a$ is an isomorphism. Let
$P_\z=\varphi(K/M,A_\z)$ and $P'=P\backslash P_\z$. Then $A_\z$
and $P_\z$ are lower-dimensional sets in $A$ and $P$,
respectively, and it is obvious that condition (a) holds. By the
computation above, $\Psi_a(\xi_{\lambda,j})\bot T_a A$, so
$Im(\Psi_a)=T_aO_a\bot T_a A$. But for $a\in A'$ $\dim T_a A+\dim
T_aO_a=\dim\La+\dim\Ll=\dim\Lp=\dim T_aP$, so $T_aP=T_a A\oplus
T_aO_a$ orthogonally. This means conditions (b) and (d) hold.
Similar to the proof of Theorem \ref{T:linear}, the dimension
condition (c) also holds. This proves that the system
$(K,\sigma,P,p(x)dx,A,da)$ is a generalized random matrix
ensemble. Since $|\det((d\pi)_e|_\Ll)|=1$, by Formula
\eqref{E:main}, we have {\small\begin{align*} J(a)
=&|\det\Psi_a|\\
=&\prod_{\lambda\in\Sigma^+}\prod_{j=1}^{\beta_{\lambda}}|\Psi_a(\xi_{\lambda,j})|\\
=&2^{\dim\Ll}\prod_{\lambda\in\Sigma^+}\left(\left|
\sinh\frac{\lambda(\eta)}{2}\,\right|
\sqrt{\cosh{\lambda(\eta)}}\right)^{\beta_{\lambda}}.
\end{align*}}
\end{proof}

The above theorem tells us that the generalized eigenvalue
distribution $d\nu$ is given by {\small\begin{equation}
d\nu(a)=2^{\dim\Ll}p(a)\prod_{\lambda\in\Sigma^+}\left(\left|
\sinh\frac{\lambda(\eta)}{2}\,\right|
\sqrt{\cosh{\lambda(\eta)}}\right)^{\beta_{\lambda}}da,
\end{equation}}
where $\eta=\log a$.

\begin{corollary}\label{C:nonlinearInt}
Let $f\in C^\infty(P)$ satisfies $f\geq0$ or $f\in L^1(P,p(x)dx)$.
Then we have the following integration formula
{\small\begin{equation}\label{E:nonlinearInt} \int_P f(x)p(x)
dx=\frac{1}{|W|}\int_A\left(\int_{K/M}f(\sigma_k(a))d\mu([k])\right)
d\nu(a).
\end{equation}}
\end{corollary}

\begin{proof}
Similar to the proof of Corollary \ref{C:linearInt}, it is
sufficient to check the covering condition (e) hold with covering
sheet $|W|$. But we notice that $\exp|_\Lp:\Lp\rightarrow P$ is a
diffeomorphism, and $\exp|_\Lp(\Lp_\z)=P_\z$,
$\exp|_\Lp(\Lp')=P'$. So the proof reduces to that of Corollary
\ref{C:linearInt}.
\end{proof}

Note that the space $G/K$ is a Riemannian symmetric space of
noncompact type, and the map $\phi:G/K\rightarrow P$ defined by
$\phi([g])=g\Theta(g)^{-1}$ is a diffeomorphism (see \cite{AW}).
So Corollary \ref{C:nonlinearInt} can be viewed as an integration
formula for symmetric space of noncompact type. Now we make it
precisely. Under the identification $\Lp\cong T_{[e]}(G/K)$, the
inner product $B|_{\Lp}$ induces a $G$-invariant Riemannian
structure on $G/K$, and then induces a $G$-invariant measure
$d\mu_1$ on $G/K$. Then we have

\begin{corollary}\label{C:symmetricInt1}
Let $f\in C^\infty(G/K)$ satisfies $f\geq0$ or $f\in
L^1(G/K,d\mu_1)$. Then
{\small\begin{equation}\label{E:symmetricInt1}
\int_{G/K}f([g])d\mu_1([g])
=\frac{1}{|W|}\int_A\left(\int_{K/M}f([ka])d\mu([k])\right)\delta(a)
da,
\end{equation}}
where
{\small$$\delta(a)=\prod_{\lambda\in\Sigma^+}\left|\sinh\lambda(\eta)\right|^{\beta_{\lambda}},$$}
here $\eta=\log a$.
\end{corollary}

\begin{proof}
First we compute the expression $|\det(d\phi)_{[a]}|$ for $a\in
A$. Choose an orthonormal basis $\eta_1,\cdots,\eta_{\dim\La}$ of
$\La$. Then the set
{\small$$\left\{\frac{d}{dt}\Big|_{t=0}[ae^{t\eta_j}]:1\leq
j\leq\dim\La\right\}\bigcup
\left\{\frac{d}{dt}\Big|_{t=0}[ae^{t\zeta_{\lambda,j}}]:\lambda\in\Sigma^+,
1\leq j\leq\beta_\lambda\right\}$$} is an orthonormal basis of
$T_{[a]}(G/K)$. It is easy to show that
{\small$$(d\phi)_{[a]}\left(\frac{d}{dt}\Big|_{t=0}[ae^{t\eta_j}]\right)=(dl_{a^2})2\eta_j,$$
$$(d\phi)_{[a]}\left(\frac{d}{dt}\Big|_{t=0}[ae^{t\zeta_{\lambda,j}}]\right)
=(dl_{a^2})2(e^{-\lambda(\eta)}\gamma_{\lambda,j}-e^{\lambda(\eta)}\theta\gamma_{\lambda,j}),$$}
where $\eta=\log a$. Then {\small\begin{align*}
|\det(d\phi)_{[a]}|
=&\prod_{j=1}^{\dim\La}|2\eta_j|\prod_{\lambda\in\Sigma^+}\prod_{j=1}^{\beta_\lambda}
|2(e^{-\lambda(\eta)}\gamma_{\lambda,j}-e^{\lambda(\eta)}\theta\gamma_{\lambda,j})|\\
=&2^{\dim\Lp}\prod_{\lambda\in\Sigma^+}\left(\sqrt{\cosh2\lambda(\eta)}\right)^{\beta_{\lambda}}.
\end{align*}}
To fulfill the proof of the corollary, we define some auxiliary
maps. Let $\psi:K/M\times A\rightarrow P$ be
$\psi([k],a)=ka^2k^{-1}$, $\rho:K/M\times A\rightarrow G/K$ be
$\rho([k],a)=[ka]$, and $sq:A\rightarrow A$ be $sq(a)=a^2$. Then
one can easily check that
$\psi=\phi\circ\rho=\varphi\circ(id\times sq),$ form which we can
easily get
\begin{align*}
\rho^*(d\mu_1)
=&|\det(d\phi)_{[a]}|^{-1}2^{\dim\La}J(a^2)d\mu da\\
=&\prod_{\lambda\in\Sigma^+}\left|\sinh\lambda(\eta)\right|^{\beta_{\lambda}}d\mu
da.
\end{align*}
Since $\rho=(\phi)^{-1}\circ\varphi\circ(id\times sq)$ is a $|W|$
sheeted covering map, by Proposition 3.1 in \cite{AWY}, we get the
desired integration formula \eqref{E:symmetricInt1}.
\end{proof}

\begin{remark}
Formula \eqref{E:symmetricInt1} has appeared in Helgason \cite{He}
(Chapter 1, Theorem 5.8).
\end{remark}

The following two examples are nonlinear versions of Example
\ref{Ex:GL(n,K)linear} and \ref{Ex:SO(m,n)linear} in the previous
section.

\begin{example}\label{Ex:GL(n,K)nonlinear}
Let $G=GL(n,\KK)$, where $\KK$ is $\RR$, $\CC$, or $\HH$. Then $G$
is real reductive. We choose the Cartan involution of
$\Lg=\gl(n,\KK)$ as $\theta(\xi)=-\xi^*$, then the corresponding
global Cartan involution of $GL(n,\KK)$ is $\Theta(g)=(g^*)^{-1}$
(see Example \ref{Ex:GL(n,K)linear}). Recall that The
corresponding Cartan decomposition of $\gl(n,\KK)$ is
$\Lk=\{\xi\in\Lg\Ll(n,\mathbb{K}):\xi^*=-\xi\}$,
$\Lp=\{\xi\in\Lg\Ll(n,\mathbb{K}):\xi^*=\xi\}$. The space
$\La=\{\eta=\diag(x_1,\cdots,x_n):x_k\in\mathbb{R}\}$ is a maximal
abelian subspace of $\Lp$ for each of the three cases, and the
subgroup $A=\exp(\La)=\{a=\diag(a_1,\cdots,a_n):a_k>0\}$. For
$\KK=\RR$, the maximal compact subgroup $K=\{g\in
GL(n,\RR):(g^t)^{-1}=g\}$ is $O(n)$. The closed submanifold
$P=\exp(\Lp)$, which is the identity component of $\{g\in
GL(n,\RR):g^t=g\}$, is the set of all real symmetric
positive-definite matrices. For $\KK=\CC$, the maximal compact
subgroup $K=\{g\in GL(n,\CC):(g^*)^{-1}=g\}$ is $U(n)$. Now the
closed submanifold $P$ is the set of all complex Hermitian
positive-definite matrices. For the case that $\KK=\HH$, the
maximal compact subgroup $K=\{g\in GL(n,\HH):(g^*)^{-1}=g\}$ is
$Sp(n)$. Now the closed submanifold $P$, which is the identity
component of $\{g\in GL(n,\HH):g^*=g\}$, is the set of all
quaternion self-adjoint positive-definite matrices. Recall that we
can choose the positive restricted root system as
$\Sigma^+=\{\ee_r-\ee_s:1\leq r<s\leq n\}$ for each case, and
$\beta_{\ee_r-\ee_s}=\beta$, where $\beta=1, 2$, or $4$ when
$\KK=\RR, \CC$, or $\HH$. For $a=\diag(a_1,\cdots,a_n)\in A$, let
$\eta=\log a=\diag(x_1,\cdots,x_n)\in\La$, where $x_k=\log a_k$.
Then by Theorem \ref{T:nonlinear},
{\small\begin{align}\label{E:J:GL(n,K)nonlinear}
J(a)=&2^{\frac{\beta n(n-1)}{2}}\prod_{1\leq r<s\leq n}
\left|\sinh\frac{x_r-x_s}{2}\right|^\beta(\cosh(x_r-x_s))^\frac{\beta}{2}\\
=&2^{-\frac{\beta n(n-1)}{4}}\prod_{1\leq r<s\leq
n}|a_r-a_s|^\beta(a_r^{-2}+a_s^{-2})^\frac{\beta}{2}.\notag
\end{align}}
Let $p(x)$ be a $K$-invariant positive smooth function on $P$,
then the density function $\mathcal{P}(a)=p(a)J(a)$ for the
nonlinear noncompact ensemble $(K,\sigma,P,p(x)dx,A,da)$ is
determined by Formula \eqref{E:J:GL(n,K)nonlinear}. For some
particular choice of $p(x)$, these kinds of ensembles are called
the new transfer matrix ensembles in \cite{CM}. But their density
functions were not derived there.\qed
\end{example}

\begin{example}\label{Ex:SO(m,n)nonlinear}
Let $G=O(m,n)_0$, $U(m,n)$, or $Sp(m,n)$. Without loss of
generality, we assume $m\geq n$. The Lie algebra for each of the
three groups has been given in Example \ref{Ex:SO(m,n)linear}. We
choose the Cartan involution of the Lie algebra as
$\theta(\xi)=-\xi^*$ with the corresponding global Cartan
involution of the group as $\Theta(g)=(g^*)^{-1}$. Then the
corresponding Cartan decomposition is $ \Lk=\left\{{\SMALL
\begin{pmatrix}
A & 0\\
0 & D
\end{pmatrix}}\right\},
\Lp=\left\{{\SMALL
\begin{pmatrix}
0 & B\\
B^* & 0
\end{pmatrix}}\right\},
$ and the corresponding maximal compact subgroup $K=S(O(m)\times
O(n))$, $U(m)\times U(n)$, or $Sp(m)\times Sp(n)$, respectively.
It is easy to show that the subgroup $A=\exp(\La)$ corresponding
to the maximal abelian subspace \eqref{E:a} of $\Lp$ is
{\small\begin{align*}
A=&\Big\{a=\sum_{k=1}^na_k(E_{m-k+1,m-k+1}+E_{m+k,m+k})\\
&+\sum_{k=1}^n\pm\sqrt{a_k^2-1}(E_{m-k+1,m+k}+E_{m+k,m-k+1}):a_k\geq1\Big\}.
\end{align*}}
In fact, under the exponential map $\exp:\La\rightarrow A$,
$a_k=\cosh x_k$. The closed submanifold $P=\exp(\Lp)$ is the
identity component of the set $\{g\in G:g^*=g\}$ for each case.
Recall that we can choose the positive restricted root system as
\eqref{E:root:O(m,n)}, and we have $\beta_{\ee_r\pm\ee_s}=\beta$,
$\beta_{\ee_r}=\beta(m-n)$, $\beta_{\ee_r}=\beta-1$, where
$\beta=1,2$, or $4$ when $G=O(m,n)_0$, $U(m,n)$, or $Sp(m,n)$. By
Theorem \ref{T:nonlinear}, the factor $J(a)$ is given by
{\small\begin{align}\label{E:(m,n)nonlinear} J(a)=&2^{n(\beta
m-1)}\prod_{1\leq r<s\leq n}
\left|\sinh\frac{x_r+x_s}{2}\sinh\frac{x_r-x_s}{2}\right|^\beta
(\cosh(x_r+x_s)\cosh(x_r-x_s))^{\frac{\beta}{2}}\notag\\
&\prod_{r=1}^n\left|\sinh\frac{x_r}{2}\right|^{\beta(m-n)}(\cosh
x_r)^{\frac{\beta(m-n)}{2}}
|\sinh x_r|^{\beta-1}(\cosh2x_r)^{\frac{\beta-1}{2}}\notag\\
=&2^{\frac{n(\beta(m+1)-2)}{2}}\prod_{1\leq r<s\leq
n}|a_r-a_s|^\beta(a_r^2+a_s^2-1)^\frac{\beta}{2}\\
&\prod_{r=1}^n\big((a_r^2-1)(2a_r^2-1)\big)^{\frac{\beta-1}{2}}
(a_r(a_r-1))^{\frac{\beta(m-n)}{2}}.\notag
\end{align}}
Let $p(x)$ be a $K$-invariant positive smooth function on $P$,
then the density function $\mathcal{P}(a)=p(a)J(a)$ for the
nonlinear noncompact ensemble $(K,\sigma,P,p(x)dx,A,da)$ is given
by \eqref{E:(m,n)nonlinear}. This may be viewed as nonlinear
noncompact orthogonal, unitary, and symplectic ensembles with
parameter $(m,n)$. \qed
\end{example}

\vskip 0.3cm The three classes of the transfer matrix ensemble are
also particular cases of the nonlinear noncompact ensemble. The
corresponding group $G=Sp(n,\RR), U(n,n)$, and $SO^*(4n)$, and the
corresponding $K=U(n), U(n)\times U(n)$, and $U(2n)$. The reader
can easily obtain the other objects and derive their joint density
functions from Theorem \ref{T:nonlinear}.

\vskip 0.5cm
\section{Compact ensembles}

\vskip 0.5cm Consider a connected compact Lie group $G$ with Lie
algebra $\Lg$. Suppose $\Theta$ is a global involutive
automorphism of $G$ with the induced involution $\theta=d\Theta$
of $\Lg$. Let $K=\{g\in G:\Theta(g)=g\}$, and let $\Lk$ and $\Lp$
be the eigenspaces of $\theta$ with eigenvalue $1$ and $-1$,
respectively. Then $\Lg=\Lk\oplus\Lp$, and we have
$[\Lk,\Lk]\subset\Lk,\; [\Lk,\Lp]\subset\Lp,\;
[\Lp,\Lp]\subset\Lk.$ Let $P=\exp(\Lp)$, then $P$ is invariant
under the conjugate action of $K$. It was proved in \cite{AW} that
$P$ is a closed submanifold of $G$ satisfies $T_eP=\Lp$, which is
just the identity component of the set $\{g\in
G:\Theta(g)=g^{-1}\}$, and we have $G=KP$. Let $G_{\CC}$ be the
complexification of $G$ with Lie algebra $\Lg_{\CC}$, then the
real Lie algebra $\Lg_*=\Lk\oplus \Lp_*$ is a real form of
$\Lg_{\CC}$, where $\Lp_*=i\Lp$. Let $G_*$ be the subgroup of
$G_{\CC}$ with Lie algebra $\Lg_*$ such that $G\cap G_*=K$, then
$G_*$ is reductive, and the direct sum $\Lg_*=\Lk\oplus \Lp_*$ is
just the Cartan decomposition of $\Lg_*$. Note that given $G_*$,
we can recover the groups $G_{\CC}$ and $G$, since $G_{\CC}$ is a
(connected) complexification of $G_*$, and $G$ is a maximal
compact group of $G_{\CC}$. Let $\La$ be a maximal abelian
subspace of $\Lp$, $A$ be the connected subgroup with Lie algebra
$\La$, which is a torus of $G$. Since $K$ is a maximal compact
subgroup of $G_*$ with Lie algebra $\Lk$ and $i\La$ is a maximal
abelian subspace of $i\Lp$, we have $i\Lp=\bigcup_{k\in
K}\Ad(k)i\La$. So $\Lp=\bigcup_{k\in K}\Ad(k)\La$, and then
$P=\bigcup_{k\in K}kAk^{-1}$. We denote $\sigma_k(p)=kpk^{-1}$ for
$k\in K$ and $p\in P$. Let $M=\{k\in K:\sigma_k(a)=a, \forall a\in
A\}$. Then $M$ is a closed subgroup of $K$ with lie algebra
$\Lm=\{\xi\in \Lk:[\xi,\eta]=0,\forall \eta\in\La\}$. The Lie
algebra $\Lg_*=\Lk\oplus i\Lp$ has the restricted root space
decomposition
$\Lg_*=(\Lg_*)_0\oplus\bigoplus_{\lambda_*\in\Sigma_*}(\Lg_*)_{\lambda_*}$,
where $\Sigma_*\subset i\La^*$ is the restricted root system of
$\Lg_*$. Define $\lambda=i\lambda_*$ and $\Sigma=i\Sigma_*$, then
$\Sigma=\{\lambda:\lambda_*\in\Sigma_*\}\subset\La^*$. Let
$\Sigma^+\subset\Sigma$ be the set of positive restricted roots.
As in \S 2, we can write $\Lk=\Lm\oplus\Ll$ and $\Lp=\La\oplus\Lb$
orthogonally. We also choose an orthogonal basis
$\{\gamma_{\lambda,1},\cdots,\gamma_{\lambda,\beta_\lambda}\}$ of
$(\Lg_*)_{\lambda_*}$ for each $\lambda\in\Sigma^+$ with
$|\gamma_{\lambda,j}|=\frac{\sqrt{2}}{2}$, and let
$\xi_{\lambda,j}=\gamma_{\lambda,j}+\theta \gamma_{\lambda,j}$,
$\zeta_{\lambda,j}=i(\gamma_{\lambda,j}-\theta
\gamma_{\lambda,j})$. Then $$\{\xi_{\lambda,j}:\lambda\in\Sigma^+,
 j=1,\cdots,\beta_\lambda\}\subset\Ll$$ is an orthonormal basis
for $\Ll$, and $$\{\zeta_{\lambda,j}:\lambda\in\Sigma^+,
j=1,\cdots,\beta_\lambda\}\subset\Lb$$ is an orthonormal basis for
$\Lb$. And we have
$\dim\Ll=\dim\Lb=\sum_{\lambda\in\Sigma^+}\beta_\lambda.$ Let
$\Ld$ be a maximal abelian subalgebra of $\Lm$, then
$\Lt=\La\oplus\Ld$ is a maximal abelian subalgebra of $\Lg$. Let
$T$ be the maximal tours with Lie algebra $\Lt$, and let
$\Delta\subset\Lt^*$ be the corresponding root system. Then
$\Sigma=\{\alpha|_\La:\alpha\in\Delta, \alpha|_\La\neq0\}$. Hence
for each $\lambda\in\Sigma$, there exists a character
$\vartheta_\lambda$ of $A$ satisfies
$\vartheta_\lambda(e^\eta)=e^{i\lambda(\eta)}, \forall\eta\in\La$.
Furthermore, we have
$\Lt\oplus\left(\Lg\bigcap\bigoplus_{\alpha|_\La=0}\Lg_{i\alpha}\right)=\La\oplus\Lm$,
where $\Lg_{i\alpha}$ is the corresponding root space in $\Lg_\CC$
(see \cite{Kn}, Formula (6.48c)).

\begin{remark}\label{R:G/K}
The global involution $\Theta$ exists if and only if $G/K$ can be
endowed with a Riemannian structure such that it is a Riemannian
symmetric space.
\end{remark}

Now we consider the action of $K$ on $P$ by
$\sigma_k(p)=kpk^{-1}$. The nondegenerate bilinear form $B_*$ on
$\Lg_*$ induces a bi-invariant Riemannian structure on $G$, such
that the linear subspaces $\La, \Lm, \mathbb{R}\xi_{\lambda,j},
\mathbb{R}\zeta_{\lambda,j}$ of $\Lg$ are mutually orthogonal.
Similar to the previous section, it induces the Riemannian
measures $dx$ and $da$ on $P$ and $A$, and $dx$ is $K$-invariant.
Since $P$ is compact, we can normalize the Riemannian structure on
$G$ such that $dx$ is a probability measure. Choose a
$K$-invariant positive smooth function $p(x)$ on $P$. As before,
the inner product $-B_*|_\Ll$ induce a $K$-invariant Riemannian
structure and a $K$-invariant Riemannian measure $d\mu$ on $K/M$.
Similar to the previous section, we define the map
$\varphi:K/M\times A\rightarrow P$ by
$\varphi([k],a)=\sigma_k(a)$. Then we can construct the
\emph{compact ensemble} with integration manifold $P$ and
eigenvalue manifold $A$ as follows.

\begin{theorem}\label{T:compact}
Let the objects be as above. Then the system
$(K,\sigma,P,p(x)dx,A,da)$ is a generalized random matrix
ensemble. Its generalized joint density function
$\mathcal{P}(a)=p(a)J(a)$ is given by
{\small\begin{equation}\label{E:compact}
J(a)=2^{\dim\Ll}\prod_{\lambda\in\Sigma^+}
\left|\sin\frac{\lambda(\eta)}{2}\right|^{\beta_\lambda},
\end{equation}}
where $\eta\in\La$ such that $e^\eta=a$.
\end{theorem}

\begin{proof}
Similar to the proof of Theorem \ref{T:nonlinear}, for $a\in A$,
we have {\Small\begin{align*} \Psi_a(\xi_{\lambda,j})
=&\frac{d}{dt}\Big|_{t=0}e^{t\xi_{\lambda,j}}ae^{-t\xi_{\lambda,j}}\\
=&(dl_a)\frac{d}{dt}\Big|_{t=0}e^{t\Ad(a^{-1})\xi_{\lambda,j}}e^{-t\xi_{\lambda,j}}\\
=&(dl_a)\left(\Ad(a^{-1})\xi_{\lambda,j}-\xi_{\lambda,j}\right)\\
=&(dl_a)\left(e^{-\ad\eta}\xi_{\lambda,j}-\xi_{\lambda,j}\right)\\
=&(dl_a)\left(e^{-\lambda_*(\eta)}\gamma_{\lambda,j}+e^{\lambda_*(\eta)}\theta\gamma_{\lambda,j}-\xi_{\lambda,j}\right)\\
=&(dl_a)\left(e^{i\lambda(\eta)}\gamma_{\lambda,j}+e^{-i\lambda(\eta)}\theta\gamma_{\lambda,j}-\xi_{\lambda,j}\right)\\
=&(dl_a)\big(\sin\lambda(\eta)\zeta_{\lambda,j}+
(\cos\lambda(\eta)-1)\xi_{\lambda,j}\big).
\end{align*}}
So {\Small\begin{align*} |\Psi_a(\xi_{\lambda,j})|
=&|\sin\lambda(\eta)\zeta_{\lambda,j}+
(\cos\lambda(\eta)-1)\xi_{\lambda,j}|\\
=&\sqrt{\sin^2\lambda(\eta)+(\cos\lambda(\eta)-1)^2}\\
=&2\left|\sin\frac{\lambda(\eta)}{2}\,\right|.
\end{align*}}
Let $A'=\{a\in
A:\vartheta_\lambda(a)\neq1,\forall\lambda\in\Sigma\}$. For
$a=e^\eta\in A'$,
$e^{i\lambda(\eta)}=\vartheta_\lambda(e^\eta)\neq1$. This implies
$\lambda(\eta)\neq2k\pi$, and hence
$|\Psi_a(\xi_{\lambda,j})|\neq0$. This means that for $a\in A'$,
$\Psi_a$ is an isomorphism. Let $A_\z=A\backslash A'$,
$P_\z=\varphi(K/M,A_\z)$, and $P'=P\backslash P_\z$. Then similar
to the proof of Theorem \ref{T:nonlinear}, one can easily check
that the conditions (a), (b), (c), and (d) hold. So the system
$(K,\sigma,P,dx,A,da)$ is a generalized random matrix ensemble.
And then {\small$$J(a)=2^{\dim\Ll}\prod_{\lambda\in\Sigma^+}
\left|\sin\frac{\lambda(\eta)}{2}\right|^{\beta_\lambda}.$$}
\end{proof}

By the above theorem, the generalized eigenvalue distribution
$d\nu$ is given by {\small\begin{equation}\label{E:compactmeasure}
d\nu(a)=2^{\dim\Ll}p(a)\prod_{\lambda\in\Sigma^+}
\left|\sin\frac{\lambda(\eta)}{2}\right|^{\beta_\lambda}da,
\end{equation}}
where $\eta\in\La$ satisfies $e^\eta=a$. Formula
\eqref{E:compactmeasure} has been obtained, when $p\equiv1$ and
omitting the constant $2^{\dim\Ll}$, by Due\~{n}ez \cite{Du} using
an integration formula associated with the $KAK$ decomposition of
compact groups. Here we recover it from Formula \eqref{E:main}
directly.

\begin{corollary}\label{C:compactInt}
Let $f\in C^\infty(P)$ satisfies $f\geq0$ or $f\in L^1(P,p(x)dx)$.
Then we have the following integration formula
{\small\begin{equation}\label{E:compactInt} \int_P f(x)
p(x)dx=\frac{1}{|W|}\int_A\left(\int_{K/M}f(\sigma_k(a))d\mu([k])\right)
d\nu(a),
\end{equation}}
where $W$ is the Weyl group of the restricted root system
$\Sigma$.
\end{corollary}

\begin{proof}
We will prove that for each $a\in A'$, the isotropic subgroup
$K_a=M$ and $|O_a\cap A'|=|W|$. Then by Corollary 3.6 in
\cite{AWY}, the covering condition (e) holds and the covering
sheet is $|W|$. By Formula \eqref{E:main2}, we get the desired
integration formulae.

\vskip 0.3cm Now we prove $K_a=M$ for $a\in A'$. First we consider
the group $Z_G(a)$. It is obvious that $MA\subset Z_G(a)$. If
$\xi\in\Lg$ lies in the Lie algebra of $Z_G(a)$, then
$\Ad_a(\xi)=\xi$, that is,
$\xi\in\Lt\oplus\left(\Lg\bigcap\bigoplus_{\vartheta_\alpha(a)=1}\Lg_{i\alpha}\right)$.
But by the definition of $A'$, $\alpha|_\La\neq0$ implies
$\vartheta_\alpha(a)\neq1$, so in fact
$\bigoplus_{\vartheta_\alpha(a)=1}\Lg_{i\alpha}=\bigoplus_{\alpha|_\La=0}\Lg_{i\alpha}$.
Hence
$\xi\in\Lt\oplus\left(\Lg\bigcap\bigoplus_{\alpha|_\La=0}\Lg_{i\alpha}\right)=\La\oplus\Lm$.
This implies that the Lie algebra of $Z_G(a)$ is $\La\oplus\Lm$.
We claim that $Z_G(a)$ is connected. In fact, let $g\in Z_G(a)$,
then the closed subgroup generated by $a$ and $g$ is a closed
abelian subgroup of $G$, hence has a generator $h$. Let $T_1$ be a
maximal torus of $G$ containing $h$, then $a,g\in T_1$. This
implies $g\in T_1\subset Z_G(a)$, thus there is a continuous path
in $T_1\subset Z_G(a)$ connecting $g$ and $e$. In a word, $Z_G(a)$
is the connected subgroup of $G$ with Lie algebra $\La\oplus\Lm$.
But $MA\subset Z_G(a)$, so in fact $Z_G(a)=MA$. (This also shows
$MA$ is connected.) Hence $K_a=Z_K(a)=Z_G(a)\cap K=MA\cap K=M$.
Here the equality $MA\cap K=M$ can be shown by an easy argument.
Next we show that $|O_a\cap A'|=|W|$ for each $a\in A'$. By the
definition of $A'$, $Z_\Lp(a):=\{\xi\in\Lp:\Ad_a(\xi)=\xi\}=\La,
\forall a\in A'$. If some $k\in K$ such that
$\sigma_k(a)=kak^{-1}\in A'$, then
$\Ad_k(\La)=\Ad_k(Z_\Lp(a))=Z_\Lp(\sigma_k(a))=\La$, that is,
$k\in N_K(\La)$. Hence $|O_a\cap
A'|=[N_K(\La):Z_K(a)]=[N_K(\La):M]=|W|$. This complete the proof
of the corollary.
\end{proof}

As we have pointed out in Remark \ref{R:G/K}, the space $G/K$ is a
Riemannian symmetric space of compact type. The map
$\phi:G/K\rightarrow P$ defined by $\phi([g])=g\Theta(g)^{-1}$ is
a diffeomorphism (see \cite{AW}). So similar to Corollary
\ref{C:symmetricInt1}, we can derive an integration formula for
symmetric space of compact type. Let $\Gamma=A\cap K$, which is a
finite group. We define the $G$-invariant measure $d\mu_1$ on
$G/K$ as in Corollary \ref{C:symmetricInt1}.

\begin{corollary}\label{C:symmetricInt2}
Under the above conditions, we have
{\small\begin{equation}\label{E:symmetricInt2}
\int_{G/K}f([g])d\mu_1([g])
=\frac{1}{|\Gamma||W|}\int_A\left(\int_{K/M}f([ka])d\mu([k])\right)\delta(a)
da,
\end{equation}}
where
$$\delta(a)=\prod_{\lambda\in\Sigma^+}
\left|\sin\lambda(\eta)\right|^{\beta_\lambda},$$ here
$\eta\in\La$ is chosen such that $e^\eta=a$.
\end{corollary}

\begin{proof}
If we define the twisted conjugate action of $G$ on $P$ by
$\tau_g(p)=gp\Theta(g)^{-1}$ (note that $\tau_k=\sigma_k$ for
$k\in K$), then it is easy to show that the measure $dx$ is
$G$-invariant, and $(\phi^{-1})^*(d\mu_1)=2^{-\dim\Lp}dx$. As in
Corollary \ref{C:symmetricInt1}, we define the maps
$\psi:K/M\times A\rightarrow P$ by $\psi([k],a)=ka^2k^{-1}$,
$\rho:K/M\times A\rightarrow G/K$ by $\rho([k],a)=[ka]$, and
$sq:A\rightarrow A$ by $sq(a)=a^2$. We also have
$\psi=\phi\circ\rho=\varphi\circ(id\times sq)$, form which one can
easily get
$$
\rho^*(d\mu_1)=\prod_{\lambda\in\Sigma^+}
\left|\sin\lambda(\eta)\right|^{\beta_\lambda}d\mu da.
$$
Since $\Gamma=\ker(sq)$,
$\rho=(\phi)^{-1}\circ\varphi\circ(id\times sq)$ is a
$|\Gamma||W|$-sheeted covering map. By Proposition 3.1 in
\cite{AWY}, the desired integration formula
\eqref{E:symmetricInt1} is proved.
\end{proof}

\begin{remark}
Formula \eqref{E:symmetricInt2} has appeared in Helgason \cite{He}
(Chapter 1, Theorem 5.10).
\end{remark}

\begin{example}\label{Ex:circular}
Let $G_*=GL(n,\KK)$, where $\KK=\RR$, $\CC$, or $\HH$. We recover
$G_{\CC}$ and $G$ below, and see what the corresponding compact
ensemble is.

\vskip 0.3cm First consider the case $G_*=GL(n,\RR)$. Its Lie
algebra $\Lg_*=\gl(n,\RR)$. $(G_*)_{\CC}=GL(n,\CC)$ is a connected
complexification of $GL(n,\RR)$, and $G=U(n)$ is a maximal compact
subgroup of $G_{\CC}$. Now
$\Lk=\Lg\cap\Lg_*=\Lu(n)\cap\gl(n,\RR)=\Lso(n)$, and $K=G\cap
G_*=U(n)\cap GL(n,\RR)=O(n)$. In the associated Cartan
decomposition of $\Lg_*=\Lk\oplus\Lp_*$, the space
$\Lp_*=\{\xi\in\gl(n,\RR):\xi^t=\xi\}$, so
$\Lp=i\Lp_*=\{i\xi:\xi\in\gl(n,\RR),\xi^t=\xi\}$, and
$\Lg=\Lk\oplus\Lp$. In fact, the global involution
$\Theta(g)=(g^t)^{-1}$ of $G=U(n)$ is compatible with the above
scheme. One can prove that the set $\{g\in
U(n):\Theta(g)=g^{-1}\}$ of symmetric unitary matrices is
connected, so we have $P=\exp(\Lp)=\{g\in U(n):g^t=g\}$, that is,
the set of $n$-by-$n$ symmetric unitary matrices. The group
$K=O(n)$ acts on $P$ by $\sigma_k(p)=kpk^{-1}$. The space
$\La=\{\eta=\diag(ix_1,\cdots,ix_n):x_k\in\RR\}$ is a maximal
abelian subspace of $\Lp$, and the corresponding eigenvalue
manifold $A=\exp(\La)=\{a=\diag(a_1,\cdots,a_n):a_k=e^{ix_k}\}$.

\vskip 0.3cm Next we let $G_*=GL(n,\CC)$. Since $\Lg_*=\gl(n,\CC)$
has a complex structure itself,
$\Lg_{\CC}\cong\gl(n,\CC)\oplus\gl(n,\CC)$ as complex Lie algebras
(see Theorem 6.94 in \cite{Kn}). So $GL(n,\CC)\times GL(n,\CC)$ is
a complexification of $GL(n,\CC)$, if we identify $G_*=GL(n,\CC)$
with the subgroup $G'_*=\{(g,\overline{g}):g\in GL(n,\CC)\}$ of
$GL(n,\CC)\times GL(n,\CC)$. The group $G=U(n)\times U(n)$ is a
maximal compact subgroup of $G_{\CC}$. Now $K=G\cap
G'_*=\big(U(n)\times U(n)\big)\cap\{(g,\overline{g}):g\in
GL(n,\CC)\}=\{(g,\overline{g}):g\in U(n)\}\cong U(n)$, and
$\Lk=\Lg\cap\Lg'_*=\{(\xi,\overline{\xi}):\xi\in \Lu(n)\}\cong
\Lu(n)$. So in the associated Cartan decomposition of
$\Lg_*=\Lk\oplus\Lp_*$, the space
$\Lp_*\cong\{(\xi,\overline{\xi}):\xi=\xi^*\}$, so
$\Lp=i\Lp_*\cong\{(\xi,\xi^t):\xi\in \Lu(n)\}$, and
$\Lg=\Lk\oplus\Lp$. Hence $P=\exp(\Lp)=\{(p,p^t):p\in U(n)\}$,
which is differmorphic to $U(n)$. In fact, the global involution
$\Theta(g_1,g_2)=(\overline{g}_2,\overline{g}_1)$ of $G=U(n)\times
U(n)$ is compatible with the above scheme. The group
$K=\{(g,\overline{g}):g\in U(n)\}$ acts on $P=\{(p,p^t):p\in
U(n)\}$ by $\sigma_k(p)=kpk^{-1}$, that is,
$\sigma_{(g,\overline{g})}(p,p^t)=(gpg^{-1},\overline{g}p^tg^t)$.
So under the identification of $G_*$ with $G'_*$, $K=P\cong U(n)$,
and the action $\sigma$ is just the conjugate action of $U(n)$.
The space
$\La=\{(\diag(ix_1,\cdots,ix_n),\diag(ix_1,\cdots,ix_n)):x_k\in\RR\}$
is a maximal abelian subspace of $\Lp$, so under the
identification of $\Lg_*$ with $\Lg'_*$,
$\La\cong\{\eta=\diag(ix_1,\cdots,ix_n):x_k\in\RR\}$. Then the
corresponding eigenvalue manifold
$A=\exp(\La)=\{a=\diag(a_1,\cdots,a_n):a_k=e^{ix_k}\}$.

\vskip 0.3cm Now we let $G_*=GL(n,\HH)$. To see what the
complexification $GL(n,\HH)_{\CC}$ is, we expand the definition of
the quaternions. Recall that an quaternion in $\HH$ is an element
of the form $z_0+\ii z_1+\jj z_2+\kk z_3$, where $z_l\in\RR$. The
multiplication in $\HH$ is defined by the linear expansion of the
relation
\begin{equation}\label{E:quaternion}
\ii^2=\jj^2=\kk^2=-1,\;\ii\jj=-\jj\ii=\kk,\;\jj\kk=-\kk\jj=\ii,\;\kk\ii=-\ii\kk=\jj.
\end{equation}
Under this multiplication, $\HH$ is a division algebra over $\RR$.
Our expanded quaternion is the element of the form $z_0+\ii
z_1+\jj z_2+\kk z_3$, where $z_l\in\CC$. We denote the set of all
such elements by $\HC$. The multiplication in $\HC$ is defined by
the complex linear extension of the relation \eqref{E:quaternion}.
This makes $\HC$ an algebra over $\CC$, and $\HC$ is the
complexification of $\HH$. But $\HC$ is not divisible. We denote
the set of all $n$-by-$n$ matrices with entries in $\HC$ by
$\gl(n,\HC)$, and the set of all invertible elements in
$\gl(n,\HC)$ by $GL(n,\HC)$. $GL(n,\HC)$ is a Lie group with Lie
algebra $\gl(n,\HC)$. Then it is easy to see that
$\gl(n,\HC)=\gl(n,\HH)_{\CC}$, $GL(n,\HC)=GL(n,\HH)_{\CC}$. In
fact, $\gl(n,\HC)\cong\gl(2n,\CC)$ as Lie algebras over $\CC$. The
isomorphism can be defined as follows. For $\xi\in\gl(n,\HC)$, let
$\xi=\xi_0+\ii\xi_1+\jj\xi_2+\kk\xi_3$, where
$\xi_l\in\gl(n,\CC)$. Define
$$\Phi(\xi)={\small
\begin{pmatrix}
\xi_0+i\xi_1 & \xi_2+i\xi_3\\
-\xi_2+i\xi_3 & \xi_0-i\xi_1
\end{pmatrix}}
\in\gl(2n,\CC),$$ then $\Phi$ is an isomorphism. In particular, we
have $\HC\cong\gl(2,\CC)$. For
$\xi=\xi_0+\ii\xi_1+\jj\xi_2+\kk\xi_3\in\gl(n,\HC)$, define the
conjugation $\overline{\xi}$ of $\xi$ by
$\overline{\xi}=\overline{\xi}_0+\ii\overline{\xi}_1+
\jj\overline{\xi}_2+\kk\overline{\xi}_3,$ and the dual $\xi^R$ of
$\xi$ by $\xi^R=\xi_0^t-\ii\xi_1^t-\jj\xi_2^t-\kk\xi_3^t.$ Define
$\xi^*=(\overline{\xi})^R$. Denote $U(n,\HC)=\{g\in
GL(n,\HC):gg^*=I_n\},$
$\Lu(n,\HC)=\{\xi\in\gl(n,\HC):\xi+\xi^*=0\},$ then $G=U(n,\HC)$
is a maximal compact subgroup of $GL(n,\HC)$ with Lie algebra
$\Lu(n,\HC)$, and $K=G_*\cap G=Sp(n)$. Note that under the
isomorphism $\Phi$ above, $U(n,\HC)\cong U(2n)$. It is easy to
show that $\Lp=\{\xi\in\Lu(n,\HC):\xi^R=\xi\}$, and $P=\{p\in
U(n,\HC):p^R=p\}$, which is the set of self-dual unitary matrices
in $GL(n,\HC)$. In fact, the global involution
$\Theta(g)=\overline{g}$ of $G=U(n,\HC)$ is compatible with the
above scheme. The group $Sp(n)$ acts on $P$ by
$\sigma_k(p)=kpk^{-1}$. The space
$\La=\{\eta=\diag(ix_1,\cdots,ix_n):x_k\in\RR\}$ is a maximal
abelian subspace of $\Lp$, and the corresponding eigenvalue
manifold $A=\exp(\La)=\{a=\diag(a_1,\cdots,a_n):a_k=e^{ix_k}\}$.

\vskip 0.3cm We can choose the set of positive restricted roots as
$\Sigma^+=\{\ee_r-\ee_s:1\leq r<s\leq n\}$ for each case, and
$\beta_{\ee_r-\ee_s}=\beta$, where $\beta=1, 2$ or $4$ when
$G_*=GL(n,\RR), GL(n,\CC)$, or $GL(n,\HH)$, respectively. Let
$p(x)$ be a $K$-invariant positive smooth function on $P$. By
Theorem \ref{T:compact}, the density function
$\mathcal{P}(a)=p(a)J(a)$ for the compact ensemble
$(G,\sigma,P,p(x)dx,A,da)$ is determined by
{\small\begin{align}\label{E:circular} J(a)=&2^{\frac{\beta
n(n-1)}{2}}\prod_{1\leq r<s\leq
n}\left|\sin\frac{x_r-x_s}{2}\right|^\beta\\
=&\prod_{1\leq r<s\leq n}|a_r-a_s|^\beta.\notag
\end{align}}
In the particular case that $p\equiv1$, the corresponding
ensembles is just the three cases of the circular ensemble.\qed
\end{example}

\begin{example}\label{Ex:Jacobi}
Let $G=SO(m+n), U(m+n)$, or $Sp(m+n)$. We choose the global
involution $\Theta$ of $G$ as $\Theta(g)=I_{m,n}gI_{m,n}$, then
$K=\{g\in G:\Theta(g)=g\}=S(O(m)\times O(n)), U(m)\times U(n)$, or
$Sp(m)\times Sp(n)$, respectively, and $P=\{g\in
G:\Theta(g)=g^{-1}\}_0=\{g\in G:(I_{m,n}g)^2=I_{m+n}\}_0$. The
induced involution $\theta=d\Theta$ of $\Lg$ is
$\theta(\xi)=I_{m,n}\xi I_{m,n}$ for $\xi\in\Lg$, and the
corresponding $ \Lk=\left\{{\SMALL
\begin{pmatrix}
A & 0\\
0 & D
\end{pmatrix}}:A+A^*=0,D+D^*=0\right\},
\Lp=\left\{{\SMALL
\begin{pmatrix}
0 & B\\
-B^* & 0
\end{pmatrix}}\right\}.
$ The space
{\small$$\La=\left\{\eta=\sum_{k=1}^nx_k(E_{m-k+1,m+k}-E_{m+k,m-k+1}):x_k\in\RR\right\}$$}
is a maximal abelian subspace of $\Lp$, and the corresponding
{\small\begin{align*}
A=&\Big\{a=\sum_{k=1}^na_k(E_{m-k+1,m-k+1}+E_{m+k,m+k})\\
&\pm\sum_{k=1}^n\sqrt{1-a_k^2}(E_{m-k+1,m+k}+E_{m+k,m-k+1}):a_k\in[-1,1]\Big\}.
\end{align*}}
In fact, under the exponential map, $a_k=\cos x_k$. It is easy to
show that $G_*$ is isomorphic to $O(m,n)_0, U(m,n)$, or $Sp(m,n)$
when $G=SO(m+n), U(m+n)$, or $Sp(m+n)$, and the positive
restricted root system $\Sigma^+$ and the associated
$\beta_\lambda$ for $\lambda\in\Sigma^+$ are the same as in
Example \ref{Ex:SO(m,n)linear} for each of the three cases. A
computation similar to that of in Example
\ref{Ex:SO(m,n)nonlinear} shows that
{\small\begin{equation}\label{E:J:Jacobi}
J(a)=2^{\frac{n(\beta(m+1)-2)}{2}}\prod_{1\leq r<s\leq
n}|a_r-a_s|^\beta\prod_{r=1}^n|1+a_r|^{\frac{\beta-1}{2}}|1-a_r|^{\frac{\beta(m-n+1)-1}{2}},
\end{equation}}
where $\beta=1, 2$, or $4$ when $G=SO(m+n), U(m+n)$, or $Sp(m+n)$,
respectively. Let $p(x)$ be a $K$-invariant positive smooth
function on $P$, then the density function
$\mathcal{P}(a)=p(a)J(a)$ for the compact ensemble
$(K,\sigma,P,p(x)dx,A,da)$ is determined by \eqref{E:J:Jacobi}. In
the particular case that $p\equiv1$, These are just three cases of
the Jacobi ensembles in Due\~{n}ez \cite{Du}.\qed
\end{example}

\vskip 0.5cm
\section{Group and algebra ensembles associated with compact groups}

\vskip 0.5cm In this section we examine the group ensemble and
algebra ensemble associated with connected compact Lie group.
First we give some general arguments.

\vskip 0.3cm Suppose $G$ is a Lie group with Lie algebra $\Lg$.
Consider the conjugate action $\sigma_g(h)=ghg^{-1}$ of $G$ on
itself and the adjoint action $\Ad_g=d\sigma_g$ of $G$ on $\Lg$.
To get the group and algebra ensembles, we need a
$\sigma$-invariant smooth measure $p(g)dg$ on $G$ and an
$\Ad$-invariant smooth measure $p(\xi)dX(\xi)$ on $\Lg$, where
$dg$ is the Haar measure on $G$ and $dX$ is the Lebesgue measure
on $\Lg$. One can easily show that such measures exist if and only
if $G$ is unimodular. In this case, we can always endow Riemannian
structures on $G$ and $\Lg$ inducing the measure $dg$ and $dX$,
respectively. To choose the zero measure subsets $X_\z$ and
$Y_\z$, we need to consider the set of singular elements in Lie
groups and Lie algebras. We denote the sets of regular elements
and singular elements in a Lie group $G$ by $G_r$ and $G_s$, and
denote the sets of regular elements and singular elements in a Lie
algebra $\Lg$ by $\Lg_r$ and $\Lg_s$.

\begin{lemma}\label{L:measurezero}
Let $M$ be a real or complex analytic manifold, $f$ an analytic
function on $M$ which is not identically zero. Then the set
$\{x\in M:f(x)=0\}$ has measure zero.
\end{lemma}

\begin{proof}
Because a complex manifold is automatically real analytic, we need
only to prove the real case. In the following we always let $f$ be
an analytic function on $M$ which is not identically zero. We
denote the zero set of $f$ by $Z$. First we suppose
$M=(-1,1)^n=\{(x_1,\cdots,x_n)\in
\mathbb{R}^n:x_j\in(-1,1),j=1,\cdots,n\}$. We prove by induction
that the zero set $Z$ of $f$ has measure zero. For $n=1$ the
conclusion is obvious true. Suppose the conclusion is true for
$n-1$. Then for the case of $n$, since $f$ is not identically
zero, the set
$$A=\{x_1\in(-1,1):f(x_1,\cdots,x_n)=0, \forall x_j\in(-1,1),
j=2,\cdots,n\}$$ is discrete, which must have measure zero in
$(-1,1)$. For $x_1\in(-1,1)\setminus A$, by the induction
hypothesis, $Z\cap(\{x_1\}\times(-1,1)^{n-1})$ has measure zero in
$\{x_1\}\times(-1,1)^{n-1}$. So by Fubini's Theorem,
{\small\begin{align*}
&\int_{(-1,1)^n}\chi_Z dx_1\cdots dx_n\\
=&\int_{x_1\in(-1,1)\setminus
A}\left(\int_{\{x_1\}\times(-1,1)^{n-1}}\chi_Z(x_1,\cdots,x_n)dx_2\cdots dx_n\right)dx_1\\
=&0,
\end{align*}}
where $\chi_Z$ is the characteristic function of $Z$. Hence $Z$
has measure zero. For the general $M$, we can choose countable may
coordinate charts $\{U_j\}_{j\in\NN}$ covering $M$ such that $U_j$
is diffeomorphic to $(-1,1)^n, \forall j\in\NN$. Then $f$ is not
identically zero on each $U_j$. Let $\nu$ be an smooth measure on
$M$, then $\nu(Z) \leq\sum_{j=1}^\infty \nu(Z\cap U_j)=0.$
\end{proof}

\begin{proposition}\label{P:measurezero}
The set of singular elements in a Lie group or a Lie algebra
always has measure zero.
\end{proposition}

\begin{proof}
Since the set of singular elements is defined to be the zero locus
of some analytic function, the proposition is obvious from the
above Lemma.
\end{proof}

Suppose $G$ is a connected compact group. Choose an
$\Ad$-invariant inner product $\langle\cdot,\cdot\rangle$ on
$\Lg$. Then it induces a bi-invariant Riemannian structure on $G$
and an $\Ad$-invariant linear Riemannian structure on $\Lg$, and
then induces a Haar measure $dg$ on $G$ and an $\Ad$-invariant
Lebesgue measure $dX$ on $\Lg$. Without loss of generality, we may
assume $dg$ is a probability measure. Let $p_{\grp}(g)$ and
$p_{\alg}(\xi)$ be $\sigma$-invariant smooth function on $G$ and
$\Ad$-invariant smooth function on $\Lg$, respectively. Let $T$ be
a maximal torus of $G$ with Lie algebra $\Lt$. Then the Riemannian
structure on $G$ also induces a Haar measure $dt$ on $T$, and the
Riemannian structure on $\Lg$ also induces a Lebesgue measure $dY$
on $\Lt$. Under the identification $\Lt^\bot\cong T_{[e]}(G/T)$,
the inner product $\langle\cdot,\cdot\rangle$ induces a
$G$-invariant Riemannian structure on $G/T$, and then induces a
$G$-invariant measure $d\mu$ on $G/T$. Since $ T=\{g\in
G:\sigma_g(t)=t, \forall t\in T\}=\{g\in G:\Ad_g(\eta)=\eta,
\forall \eta\in\Lt\}$, we can form the maps
$\varphi^\grp:G/T\times T\rightarrow G$ and
$\varphi^\alg:G/T\times\Lt\rightarrow\Lg$ by
$\varphi^\grp([g],t)=\sigma_g(t)$,
$\varphi^\alg([g],\eta)=\Ad_g(\eta)$, respectively. Let
$\Delta\subset\Lt^*$ be the root system. For $\alpha\in\Delta$,
let $\vartheta_\alpha$ be the character of $T$ defined by
$\vartheta_\alpha(e^\eta)=e^{i\alpha(\eta)}, \forall\eta\in\Lt$.

\begin{theorem}\label{T:compact-group}
Let the objects be as above. Then\\
(1) $(G,\sigma,G,p_{\grp}(g)dg,T,dt)$ is a generalized random
matrix ensemble. Its generalized joint density function
$\mathcal{P}(t)=p_{\grp}(t)J_{\grp}(t)$ is given by
{\small\begin{equation}\label{E:compact-group}
J_{\grp}(t)=\prod_{\alpha\in\Delta}|1-\vartheta_\alpha(t^{-1})|.
\end{equation}}
(2) $(G,\Ad,\Lg,p_{\alg}(\xi)dX(\xi),\Lt,dY)$ is a generalized
random matrix ensemble. Its generalized joint density function
$\mathcal{P}(\eta)=p_{\alg}(\eta)J_{\alg}(\eta)$ is given by
{\small\begin{equation}\label{E:compact-algebra}
J^\alg(\eta)=\prod_{\alpha\in\Delta}|\alpha(\eta)|.
\end{equation}}
\end{theorem}

\begin{proof}
Let $G_\z=G_s, T_\z=T\cap G_\z$. By Proposition
\ref{P:measurezero}, $G_\z$ has measure zero in $G$. Since
$T_\z=\bigcup_{\alpha\in\Delta}\ker\vartheta_\alpha$, $T_\z$ has
measure zero in $T$. Let $G'=G\setminus G_\z=G_r$ and
$T'=T\setminus T_\z$. By the theory of compact Lie groups, one can
easily show that the conditions (a), (b), (c), and (d) hold, and
then uses Formula \eqref{E:main} to prove formula
\eqref{E:compact-group}. This proves (1). (2) can be proved
similarly.
\end{proof}

\begin{corollary}\label{C:compact-groupInt}
Let $G$ be a compact group with a maximal torus $T$, and let $\Lg$
and $\Lt$ be their Lie algebras. Then we have
{\small\begin{equation}\label{E:compact-groupInt} \int_G f(g)
dg=\frac{1}{|W|}\int_T\left(\int_{G/T}f(\sigma_g(t))d\mu([g])\right)J^\grp(t)
dt,
\end{equation}
\begin{equation}\label{E:compact-algebraInt}
\int_\Lg f(\xi)
dX(\xi)=\frac{1}{|W|}\int_\Lt\left(\int_{G/T}f(\Ad_g(\eta))d\mu([g])\right)J^\alg(\eta)
dY(\eta).
\end{equation}}
\end{corollary}

\begin{proof}
Using Corollary 3.6 in \cite{AWY}, one can easily shows that for
both cases in Theorem \ref{T:compact-group}, the covering
condition (e) holds, and the covering sheet is $|W|$. So the
corollary directly from Formula \eqref{E:main2}.
\end{proof}

\begin{remark}
Formula \eqref{E:compact-groupInt} is just the Weyl integration
formula for compact Lie groups. \eqref{E:compact-algebraInt} can
be viewed as the linear version of the the Weyl integration
formula. Here we recover them from the viewpoint of generalized
random matrices.
\end{remark}

\begin{example}\label{Ex:compact}
Let $G=U(n), SO(2n+1), Sp(n)$, or $SO(2n)$. We derive the joint
density functions for the corresponding group ensemble and algebra
ensemble by deriving the factor $J^\alg(\eta)$ and $J^\grp(t)$ for
each case.

\vskip 0.3cm First we let $G=U(n)$. Then
$T=\{t=\diag(t_1,\cdots,t_n):|t_k|=1\}$ is a maximal torus of
$U(n)$ with Lie algebra
$\Lt=\{\eta=\diag(ix_1,\cdots,ix_n):x_k\in\RR\}$. The associated
root system $\Delta=\{\pm(\ee_r-\ee_s):1\leq r<s\leq n\}$, where
$\ee_r\in\Lt^*$ is defined by
$\ee_r(\diag(ix_1,\cdots,ix_n))=x_r$. Then by Theorem
\ref{T:compact-group}, for the algebra ensemble
$(U(n),\Ad,\Lu(n),p(\xi)dX(\xi),\Lt,dY)$, the factor
{\small\begin{equation} J^\alg(\eta)=\prod_{1\leq r<s\leq
n}|x_r-x_s|^2.
\end{equation}}
And for the group ensemble $(U(n),\sigma,U(n),dg,T,dt)$, the
density function {\small\begin{equation} J^\grp(t)=\prod_{1\leq
r<s\leq n}|1-e^{i(x_r-x_s)}|^2=\prod_{1\leq r<s\leq n}|t_r-t_s|^2,
\end{equation}}
where we have chosen $\eta=\diag(ix_1,\cdots,ix_n)\in\Lt$ such
that $t=e^\eta$.

\vskip 0.3cm Next we consider the case $G=SO(2n+1)$. Then the
maximal torus of $SO(2n+1)$ can be chosen as
$$T=\left\{t=\diag\left({\SMALL
\begin{pmatrix}
t_1 & -t'_1\\
t'_1 & t_1
\end{pmatrix},\cdots,
\begin{pmatrix}
t_n & -t'_n\\
t'_n & t_n
\end{pmatrix}},1
\right):t_k, t'_k\in[-1,1],t_k^2+t_k'^2=1\right\},$$ whose Lie
algebra is
$$\Lt=\left\{\eta=\diag\left({\SMALL
\begin{pmatrix}
0 & -x_1\\
x_1 & 0
\end{pmatrix},\cdots,
\begin{pmatrix}
0 & -x_n\\
x_n & 0
\end{pmatrix}},0
\right):x_k\in\RR\right\}.$$ The root system
$\Delta=\{\pm(\ee_r+\ee_s),\pm(\ee_r-\ee_s):1\leq r<s\leq
n\}\cup\{\pm\ee_r:1\leq r\leq n\}$, where $\ee_r(\eta)=x_r$. By
Theorem \ref{T:compact-group}, for the algebra ensemble
$(SO(2n+1),\Ad,\Lso(2n+1),p(\xi)dX(\xi),\Lt,dY)$, the factor
{\small\begin{equation} J^\alg(\eta)=\prod_{1\leq r<s\leq
n}|x_r+x_s|^2|x_r-x_s|^2\prod_{r=1}^n|x_r|^2=\prod_{1\leq r<s\leq
n}|x_r^2-x_s^2|^2\prod_{r=1}^n|x_r|^2.
\end{equation}}
If for $t\in T$ we choose $\eta\in\Lt$ such that $t=e^\eta$, that
is, $t_k=\cos x_k, t'_k=\sin x_k$, then the density function for
the group ensemble $(SO(2n+1),\sigma,SO(2n+1),dg,T,dt)$ is
{\small\begin{align} J^\grp(t)=&\prod_{1\leq r<s\leq
n}|1-e^{i(x_r+x_s)}|^2|1-e^{i(x_r-x_s)}|^2\prod_{r=1}^n|1-e^{ix_r}|^2\\
=&2^{n^2}\prod_{1\leq r<s\leq
n}(t_r-t_s)^2\prod_{r=1}^n(1-t_r).\notag
\end{align}}

\vskip 0.3cm Now we let $G=Sp(n)$. Then
$T=\{t=\diag(t_1,\cdots,t_n,\overline{t}_1,\cdots,\overline{t}_n):|t_k|=1\}$
is a maximal torus of $U(n)$ with Lie algebra
$\Lt=\{\eta=\diag(x_1,\cdots,x_n,-x_1,\cdots,-x_n):x_k\in\RR\}$.
The root system $\Delta=\{\pm(\ee_r+\ee_s),\pm(\ee_r-\ee_s):1\leq
r<s\leq n\}\cup\{\pm2\ee_r:1\leq r\leq n\}$, where
$\ee_r(\eta)=x_r$. So by Theorem \ref{T:compact-group}, for the
algebra ensemble $(Sp(n),\Ad,\Lsp(n),p(\xi)dX(\xi),\Lt,dY)$, the
factor {\small\begin{equation} J^\alg(\eta)=\prod_{1\leq r<s\leq
n}|x_r+x_s|^2|x_r-x_s|^2\prod_{r=1}^n|2x_r|^2=2^{2n}\prod_{1\leq
r<s\leq n}|x_r^2-x_s^2|^2\prod_{r=1}^n|x_r|^2.
\end{equation}}
For $t\in T$, choose $\eta\in\Lt$ such that $e^\eta=t$, then the
density function for the group ensemble
$(Sp(n),\sigma,Sp(n),dg,T,dt)$ is {\small\begin{align}
J^\grp(t)=&\prod_{1\leq r<s\leq n}|1-e^{i(x_r+x_s)}|^2|1-e^{i(x_r-x_s)}|^2\prod_{r=1}^n|1-e^{2ix_r}|^2\\
=&\prod_{1\leq r<s\leq
n}|t_r-t_s|^2|1-t_rt_s|^2\prod_{r=1}^n|1-t_r^2|^2.\notag
\end{align}}

\vskip 0.3cm For the last group $G=SO(2n)$, the maximal torus of
$SO(2n+1)$ can be chosen as
$$T=\left\{t=\diag\left({\SMALL
\begin{pmatrix}
t_1 & -t'_1\\
t'_1 & t_1
\end{pmatrix},\cdots,
\begin{pmatrix}
t_n & -t'_n\\
t'_n & t_n
\end{pmatrix}}
\right):t_k, t'_k\in[-1,1],t_k^2+t_k'^2=1\right\},$$ whose Lie
algebra is
$$\Lt=\left\{\eta=\diag\left({\SMALL
\begin{pmatrix}
0 & -x_1\\
x_1 & 0
\end{pmatrix},\cdots,
\begin{pmatrix}
0 & -x_n\\
x_n & 0
\end{pmatrix}}
\right):x_k\in\RR\right\}.$$ The root system
$\Delta=\{\pm(\ee_r+\ee_s),\pm(\ee_r-\ee_s):1\leq r<s\leq n\}$,
where $\ee_r(\eta)=x_r$. Then by Theorem \ref{T:compact-group},
for the algebra ensemble $(SO(2n), \Ad, \Lso(2n),$ $p(\xi)dX(\xi),
\Lt, dY)$, the factor {\small\begin{equation}
J^\alg(\eta)=\prod_{1\leq r<s\leq
n}|x_r+x_s|^2|x_r-x_s|^2=\prod_{1\leq r<s\leq n}|x_r^2-x_s^2|^2.
\end{equation}}
If for $t\in T$ we choose $\eta\in\Lt$ such that $t=e^\eta$, then
the density function for the group ensemble
$(SO(2n),\sigma,SO(2n),dg,T,dt)$ is {\small\begin{align}
J^\grp(t)=&\prod_{1\leq r<s\leq n}|1-e^{i(x_r+x_s)}|^2|1-e^{i(x_r-x_s)}|^2\\
=&2^{n(n-1)}\prod_{1\leq r<s\leq n}(t_r-t_s)^2.\notag
\end{align}}

\vskip 0.3cm Let $p_{\grp}(g)$ and $p_{\alg}(\xi)$ be
$\sigma$-invariant smooth function on $G$ and $\Ad$-invariant
smooth function on $\Lg$, then the density functions
$\mathcal{P}(t)=p_{\grp}(t)J_{\grp}(t)$ and
$\mathcal{P}(\eta)=p_{\alg}(\eta)J_{\alg}(\eta)$ for the group
ensemble $(G,\sigma,G,p_{\grp}(g)dg,T,dt)$ and the algebra
ensemble $(G,\Ad,\Lg,p_{\alg}(\xi)dX(\xi),\Lt,dY)$ are determined
by the above formulae. In the particular case that
$p_{\grp}\equiv1$, these four classes of group ensembles were
particularly interesting for number theorist, since they have
close relation with the distribution of the Riemann zeta function
and $L$-functions (see \cite{KS}). Note that when
$p_{\grp}\equiv1$, the group ensemble associated with $U(n)$ is
just the circular unitary ensemble, and for suitable choice of
$p_{\alg}$, the algebra ensemble associated with $U(n)$ is just
the Gaussian unitary ensemble up to multiplication by $i$. \qed
\end{example}

\vskip 0.5cm
\section{Group and algebra ensembles associated with complex semisimple Lie groups}

\vskip 0.5cm

Now we consider the group ensemble and the algebra ensemble
associated with a connected complex semisimple Lie group $G$ with
lie algebra $\Lg$. Let $\Lh$ be a Cartan subalgebra of $\Lg$, $H$
be the connected Lie subgroup of $G$ with lie algebra $\Lh$, which
is called a Cartan subgroup of $G$. Then $H=\{g\in
G:\sigma_g(h)=h, \forall h\in H\} =\{g\in G:\Ad_g(\eta)=\eta,
\forall \eta\in\Lh\}$. Similarly We can define the map
$\varphi^\grp:G/H\times H\rightarrow G$ and
$\varphi^\alg:G/H\times\Lh\rightarrow\Lg$ by
$\varphi^\grp([g],h)=\sigma_g(h)$,
$\varphi^\alg([g],\eta)=\Ad_g(\eta)$. Note that unlike the case
for compact Lie groups, the maps $\varphi^\grp$ and $\varphi^\alg$
are not surjective in general, but every regular element of $G$ or
$\Lg$ lies in the image of $\varphi^\grp$ or $\varphi^\alg$. Let
$\Delta=\Delta(\Lg,\Lh)$ be the root system. For each
$\alpha\in\Delta$, Let $\vartheta_\alpha$ be the restriction of
the adjoint representation of $H$ on the root space $\Lg_\alpha$.
Note that $\vartheta_\alpha(e^\eta)=e^{\alpha(\eta)}$ for
$\eta\in\Lh$. Choose a left invariant Riemannian structure on $G$
such that $\Lh$ and the root spaces $\Lg_\alpha$ are mutually
orthogonal. It induces Haar measures $dg,dh$ on $G$ and $H$, and
the associated linear Riemannian structure on $\Lg$ induces
Lebesgue measures $dX, dY$ on $\Lg$ and $\Lh$. We choose a
Riemannian structure on $G/H$ which induces a $G$-invariant
Riemannian measure $d\mu$ on $G/H$, such that the identification
$\Lh^\bot=\bigoplus_{\alpha\in\Delta}\Lg_\alpha\cong T_{[e]}(G/H)$
is isometric. To simplify to notations, we let the functions
$p_\grp\equiv1$ and $p_\alg\equiv1$.

\begin{theorem}\label{T:complex-group}
Let the objects be as above. Then\\
(1) $(G,\sigma,G,dg,H,dh)$ is a generalized random matrix
ensemble. Its generalized joint density function
$\mathcal{P}(h)=J^\grp(h)$ is given by
{\small\begin{equation}\label{E:complex-group}
J^\grp(h)=\prod_{\alpha\in\Delta}|1-\vartheta_\alpha(h^{-1})|^2.
\end{equation}}
(2) $(G,\Ad,\Lg,dX,\Lh,dY)$ is a generalized random matrix
ensemble. Its generalized joint density function
$\mathcal{P}(\eta)=J^\alg(\eta)$ is given by
{\small\begin{equation}\label{E:complex-algebra}
J^\alg(\eta)=\prod_{\alpha\in\Delta}|\alpha(\eta)|^2.
\end{equation}}
\end{theorem}

\begin{proof}
We first prove (1). Let $G_\z=G_s, H_\z=H\cap
G_\z=\bigcup_{\alpha\in\Delta}\ker\vartheta_\alpha$, then $G_\z$
and $H_\z$ has measure zero in $G$ and $H$, respectively. Let
$G'=G\setminus G_\z=G_r$ and $H'=H\setminus H_\z$. We prove the
conditions (a), (b), (c), and (d) hold. For every $g\in G'$, there
is some $g'\in G$ such that $\sigma_{g'}(g)\in H'$, so every orbit
in $G'$ intersects $H'$. Then the invariance condition (a) holds.
We denote $\Lg_1=\bigoplus_{\alpha\in\Delta}\Lg_\alpha$. For $h\in
H$, consider the map $\Psi_h:\Lg_1\rightarrow T_h O_h$,
$\Psi_h(\xi)=\frac{d}{dt}\big|_{t=0}\sigma_{\exp t\xi}(h)$. If we
identify $T_h G$ with $\Lg=T_e G$ by left translation, we have,
for $\xi\in\Lg_\alpha$, {\small\begin{align*} \Psi_h(\xi)
=&(dl_{h^{-1}})_h\frac{d}{dt}\Big|_{t=0}\exp(t\xi)h\exp(-t\xi)\\
=&\frac{d}{dt}\Big|_{t=0}\left(h^{-1}\exp(t\xi)h\right)\exp(-t\xi)\\
=&\Ad(h^{-1})(\xi)-\xi\\
=&\left(\vartheta_\alpha(h^{-1})-1\right)\xi.
\end{align*}}
If $h\in H'$, $\vartheta_\alpha(h^{-1})-1\neq0,
\forall\alpha\in\Delta$, so $\Psi_h$ is an isomorphism from
$\Lg_1$ onto $T_h O_h=\Lg_1$. Then we have $T_h G=\Lh\oplus T_h
O_h$ orthogonally, that is, the conditions (b) and (d) hold. By
Corollary 7.106 in \cite{Kn}, the identity component of $G_h$ is
$H$, $\forall h\in H'$, so the dimension condition (c) also holds.
We have shown that $\Psi_h$ acts on
$\Lg_1=\bigoplus_{\alpha\in\Delta}\Lg_\alpha$ diagonally with
eigenvalues
$\{\lambda_\alpha=\vartheta_\alpha(h^{-1})-1:\alpha\in\Delta\}$.
Each eigenspace has complex dimension 1, But what we are looking
for is the norm of the ``determinant" $|\det\Psi_h|$ of $\Psi_h$,
which was regarded as a real linear map. Note that if we view
$\mathbb{C}$ as a 2-dimensional real vector space with a basis
$(1,i)$, then multiplication by $\lambda_\alpha$ induces a linear
transformation with matrix $ {\SMALL\begin{pmatrix}
Re\lambda_\alpha & -Im\lambda_\alpha\\
Im\lambda_\alpha & Re\lambda_\alpha
\end{pmatrix}},
$ whose determinant is $|\lambda_\alpha|^2$. Note that the
identification $\Lg_1\cong T_{[e](G/H)}$ is isometric, we have
{\small\begin{align*}
J^\grp(h)=&|\det\Psi_h|\\
=&\prod_{\alpha\in\Delta}|\lambda_\alpha|^2\\
=&\prod_{\alpha\in\Delta}|\vartheta_\alpha(h^{-1})-1|^2.
\end{align*}}
This proves (1). The proof of (2) is similar but more easy. We
omit it here.
\end{proof}

\begin{corollary}\label{C:complex-groupInt}
Let $G$ be a complex semisimple Lie group with a Cartan subgroup
$H$, and let $\Lg$ and $\Lh$ be their Lie algebras. Then we have
{\small\begin{equation}\label{E:complex-groupInt} \int_G f(g)
dg=\frac{1}{|W|}\int_H\left(\int_{G/H}f(\sigma_g(h))d\mu([g])\right)J^\grp(h)
dh,
\end{equation}
\begin{equation}\label{E:complex-algebraInt}
\int_\Lg f(\xi)
dX(\xi)=\frac{1}{|W|}\int_\Lh\left(\int_{G/H}f(\Ad_g(\eta))d\mu([g])\right)J^\alg(\eta)
dY(\eta),
\end{equation}}
where $W=W(\Delta)$ is the Weyl group.
\end{corollary}

\begin{proof}
By Formula \eqref{E:main2}, it is sufficient to check the covering
condition (e) for both cases and show the covering sheet is $|W|$.
For the Lie algebra case, by the structure theory of complex
semisimple Lie group, for every $\eta\in\Lh'$, the isotropic
subgroup $G_\eta$ of $G$ associated with $\eta$ equals to $H$. It
is also known that for every $\xi\in\Lg'$, there exists some $g\in
G$ such that $\Ad_g(\xi)\in\Lh'$. Such $\Ad_g$ are labelled by
$N_G(\Lh)=\{\tau\in \mathrm{Int}(\Lg):\tau(\Lh)=\Lh\}$, that is,
if $\Ad_g$ and $\Ad_{g'}$ both send $\xi$ into $\Lh'$, then
$\Ad_g=\tau\circ\Ad_{g'}$ for some $\tau\in N_G(\Lh)$, and such
$\tau$ is unique. But it is known that $W\cong N_G(\Lh)/H$. So
every orbit in $\Lg'$ intersects $\Lh'$ at $|W|$ points. By
Corollary 3.6 in \cite{AWY}, the covering condition (e) holds for
the Lie algebra case, and the covering sheet is $|W|$.

\vskip 0.3cm Now we prove the Lie group case. We want to show that
$\forall g\in G'$, $(\varphi^{\grp})^{-1}(g)$ has $|W|$ points. By
the same reason as in the proof of Corollary 3.6 in \cite{AWY}, we
need only to show the case $g\in H'$, and the general case can be
reduced to it. Thus we let $h\in H'$, and let
$g_1,\cdots,g_{|W|}\in N_G(\Lh)$, one in each component. Then
$\{([g_i],g_i^{-1}hg_i):i=1,\cdots,|W|\}\subset(\varphi^{\grp})^{-1}(h)$.
Let $([g],h')\in(\varphi^{\grp})^{-1}(h)$, then $gh'g^{-1}=h$. But
$h$ and $h'$ are regular, their centralizer in $\Lg$ must be the
Cartan subalgebra $\Lh$. So $\Ad(g)$ fixes $\Lh$, and then $g\in
N_G(\Lh)$, so $([g],h')=([g_{i_0}],g_{i_0}^{-1}hg_{i_0})$ for some
$i_0\in\{1,\cdots,|W|\}$. Thus in fact we have
$(\varphi^{\grp})^{-1}(h)=\{([g_i],g_i^{-1}hg_i):i=1,\cdots,|W|\}$,
which has $|W|$ point. By Proposition 3.5 in \cite{AWY},
$\varphi^{\grp}$ is a $|W|$ sheeted covering map.
\end{proof}

\begin{remark}\label{R}
Notice that when we prove $\varphi^{\grp}$ is a covering map, we
make use of Proposition 3.5 in \cite{AWY} directly, ignoring
Corollary 3.6 in \cite{AWY}. In fact, the conditions of Corollary
3.6 in \cite{AWY} are not satisfied in general, that is, here the
phenomena of sudden variation of the isotropic subgroups may
happens (see Remark 3.2 in \cite{AWY}).
$G=SL(2,\mathbb{C})/\{\pm1\}$ is such an example. For details see
(\cite{Kn}, Section 7.8).
\end{remark}

\begin{remark}
Formula \eqref{E:complex-groupInt} is just Harish-Chandra's
integration formula for complex semisimple Lie groups.
\eqref{E:complex-algebraInt} is the linear version of
\eqref{E:complex-groupInt}. Here we recover them form the
viewpoint of generalized random matrices.
\end{remark}

\begin{example}\label{Ex:SL(n,C)}
Let $G=SL(n,\CC) (n\geq2)$, which is a complex simple Lie group.
{\small$$\Lh=\{\eta=\diag(x_1,\cdots,x_n):x_k\in\CC,
\sum_{k=1}^nx_k=0\}$$} is a Cartan subalgebra of the Lie algebra
$\Lg=\Lsl(n,\CC)$ of $G$. The corresponding Cartan subgroup is
{\small$$H=\{h=\diag(h_1,\cdots,h_n):h_k\in\CC,
\prod_{k=1}^nh_k=1\}.$$} The root system
$\Delta=\{\pm(\ee_r-\ee_s):1\leq r<s\leq n\},$ where
$\ee_r\in\Lh^*$ is defined by $\ee_r(\diag(x_1,\cdots,x_n))=x_r$.
So by Theorem \ref{T:complex-group}, the generalized joint density
function $\mathcal{P}(\eta)=J^\alg(\eta)$ for the algebra ensemble
$(SL(n,\CC),\Ad,\Lsl(n,\CC),dX,\Lh,dY)$ is {\small\begin{equation}
J^\alg(\eta)=\prod_{1\leq r<s\leq n}|x_r-x_s|^4.
\end{equation}}
For $h=\diag(h_1,\cdots,h_n)\in H$, choose some
$\eta=\diag(x_1,\cdots,x_n)\in\Lh$ such that $h=e^\eta$, that is,
$h_r=e^{x_r}$ for each $r$. Since
$\vartheta_\alpha(h)=\vartheta_\alpha(e^\eta)=e^{\alpha(\eta)}$
for each $\alpha\in\Delta$, by Theorem \ref{T:complex-group}, we
get the density function $\mathcal{P}(h)=J^\grp(h)$ for the group
ensemble $(SL(n,\CC),\sigma,SL(n,\CC),dg,H,dh)$ as
{\small\begin{align} J^\grp(h)=&\prod_{1\leq r<s\leq
n}|1-e^{x_r-x_s}|^2|1-e^{x_s-x_r}|^2\notag\\
=&\prod_{1\leq r<s\leq
n}|e^{x_r}-e^{x_s}|^4\prod_{r=1}^n|e^{-2(n-1)x_r}|\\
=&\prod_{1\leq r<s\leq n}|h_r-h_s|^4.\notag
\end{align}}
Note that in this example the ``eigenvalue manifolds" $H$ and
$\Lh$ are really consist of eigenvalues of the matrices in the
corresponding integration manifolds. \qed
\end{example}

\begin{example}\label{Ex:Sp(n,C)}
Let $G=Sp(n,\CC)=\{g\in SL(2n,\CC):g^tJ_{n,n}g=J_{n,n}\}$ which is
a complex simple Lie group, where $J_{n,n}={\SMALL
\begin{pmatrix}
0 & I_n\\
-I_n & 0
\end{pmatrix}}$. Its Lie algebra
$\Lg=\Lsp(n,\CC)=\{\xi\in\Lsl(2n,\CC):\xi^tJ_{n,n}+J_{n,n}\xi=0\}$.
$$\Lh=\{\eta=\diag(x_1,\cdots,x_n,-x_1,\cdots,-x_n):x_k\in\CC\}$$
is a Cartan subalgebra of $\Lsp(n,\CC)$. The corresponding Cartan
subgroup is
$$H=\{h=\diag(h_1,\cdots,h_n,h_1^{-1},\cdots,h_n^{-1}):h_k\in\CC,
h_k\neq0\}.$$ The root system
$\Delta=\{\pm(\ee_r+\ee_s),\pm(\ee_r-\ee_s):1\leq r<s\leq n\}
\cup\{\pm2\ee_r:1\leq r\leq n\}.$ By Theorem
\ref{T:complex-group}, the density function
$\mathcal{P}(\eta)=J^\alg(\eta)$ for the algebra ensemble
$(Sp(n,\CC),\Ad,\Lsp(n,\CC),dX,\Lh,dY)$ is {\small\begin{equation}
J^\alg(\eta)=2^{4n}\prod_{1\leq r<s\leq
n}|x_r^2-x_s^2|^4\prod_{r=1}^n|x_r|^4.
\end{equation}}
For $h=\diag(h_1,\cdots,h_n,h_1^{-1},\cdots,h_n^{-1})\in H$,
choose some $\eta=\diag(x_1,\cdots,x_n,$ $-x_1,\cdots,-x_n)\in\Lh$
such that $h=e^\eta$, that is, $h_r=e^{x_r}$ for each $r$. Then we
have $\vartheta_\alpha(h)=e^{\alpha(\eta)}$ for each
$\alpha\in\Delta$. By Theorem \ref{T:complex-group}, the density
function $\mathcal{P}(h)=J^\grp(h)$ for the group ensemble
$(Sp(n,\CC),\sigma,Sp(n,\CC),dg,H,dh)$ is {\small\begin{align}
J^\grp(h)=&\prod_{1\leq r<s\leq
n}|1-e^{x_r+x_s}|^2|1-e^{-(x_r+x_s)}|^2|1-e^{x_r-x_s}|^2|1-e^{x_s-x_r}|^2\notag\\
&\prod_{r=1}^n|1-e^{2x_r}|^2|1-e^{-2x_r}|^2\\
=&\prod_{1\leq r<s\leq
n}|h_r-h_s|^4|1-h_rh_s|^4\prod_{r=1}^n|1-h_r^2|^4|h_r|^{-2n(n+1)}.\notag
\end{align}}
Similar to Example \ref{Ex:SL(n,C)}, here the ``eigenvalue
manifolds" $H$ and $\Lh$ are also consist of eigenvalues of the
matrices in the corresponding integration manifolds. \qed
\end{example}

\begin{example}\label{Ex:SO(2n,C)}
Let $G=SO(2n,\CC)$. Then
$$\Lh=\left\{\eta=\diag\left({\SMALL
\begin{pmatrix}
0 & -x_1\\
x_1 & 0
\end{pmatrix},\cdots,
\begin{pmatrix}
0 & -x_n\\
x_n & 0
\end{pmatrix}}
\right):x_k\in\CC\right\}$$ is a Cartan subalgebra of
$\Lso(2n,\CC)$, the corresponding Cartan subgroup is
$$H=\left\{h=\diag\left({\SMALL
\begin{pmatrix}
h_1 & -h'_1\\
h'_1 & h_1
\end{pmatrix},\cdots,
\begin{pmatrix}
h_n & -h'_n\\
h'_n & h_n
\end{pmatrix}}
\right):h_k, h'_k\in\CC,h_k^2+h_k'^2=1\right\}.$$ A routine
computation similar to that of in Example \ref{Ex:SL(n,C)} and
\ref{Ex:Sp(n,C)} shows that {\small\begin{equation}
J^\alg(\eta)=\prod_{1\leq r<s\leq n}|x_r^2-x_s^2|^4,
\end{equation}
\begin{equation}
J^\grp(h)=2^{2n(n-1)}\prod_{1\leq r<s\leq n}|h_r-h_s|^4.
\end{equation}}\qed
\end{example}

\begin{example}\label{Ex:SO(2n+1,C)}
Let $G=SO(2n+1,\CC)$. Then
$$\Lh=\left\{\eta=\diag\left({\SMALL
\begin{pmatrix}
0 & -x_1\\
x_1 & 0
\end{pmatrix},\cdots,
\begin{pmatrix}
0 & -x_n\\
x_n & 0
\end{pmatrix}},0
\right):x_k\in\CC\right\}$$ is a Cartan subalgebra of
$\Lso(2n+1,\CC)$, and the corresponding Cartan subgroup is
$$H=\left\{h=\diag\left({\SMALL
\begin{pmatrix}
h_1 & -h'_1\\
h'_1 & h_1
\end{pmatrix},\cdots,
\begin{pmatrix}
h_n & -h'_n\\
h'_n & h_n
\end{pmatrix}},1
\right):h_k, h'_k\in\CC,h_k^2+h_k'^2=1\right\}.$$ Then we can
derive that {\small\begin{equation} J^\alg(\eta)=\prod_{1\leq
r<s\leq n}|x_r^2-x_s^2|^4\prod_{r=1}^n|x_r|^4,
\end{equation}
\begin{equation}
J^\grp(h)=2^{2n^2}\prod_{1\leq r<s\leq
n}|h_r-h_s|^4\prod_{r=1}^n|1-h_r|^2.
\end{equation}}\qed
\end{example}

\vskip 0.5cm
\section{Pseudo-group and pseudo-algebra ensembles}

\vskip 0.5cm

In this section we consider the pseudo-group ensemble and the
pseudo-algebra ensemble associated with a real reductive group.
Strictly speaking, they are not generalized ensembles, since the
integration manifolds may have singularities. But this doesn't
matter, since integration manifold is the closure of an open
submanifold of a real reductive group or a real reductive Lie
algebra, whose boundary has measure zero. Let $G$ be a real
reductive group with lie algebra $\Lg$. Let $\theta$ be a Cartan
involution of $\Lg$, and let $\Lh_1,\cdots,\Lh_m$ be a maximal set
of mutually nonconjugate $\theta$ stable Cartan subalgebras of
$\Lg$. The corresponding Cartan subgroups of $G$ are
$H_1=Z_G(\Lh_1),\cdots,H_m=Z_G(\Lh_m)$. Let $G'=G_r$,
$H'_j=H_j\cap G'$, and let $\Lg'=\Lg_r$, $\Lh'_j=\Lh_j\cap\Lg'$.
Then it is known that $G'=\bigsqcup_{j=1}^m\bigcup_{g\in
G}gH'_jg^{-1}$, $\Lg'=\bigsqcup_{j=1}^m\bigcup_{g\in
G}\Ad(g)(\Lh'_j)$. Denote $G'_j=\bigcup_{g\in G}gH'_jg^{-1}$,
$\Lg'_j=\bigcup_{g\in G}\Ad(g)(\Lh'_j)$. Then each $\Lg'_j$ is an
open set in $\Lg$, and each $G'_j$ is an open set in $G$. Let
$G_j=\overline{G'_j}$, $\Lg_j=\overline{\Lg'_j}$. It is easy to
show that $\{g\in G:\sigma_g(h)=h,\forall h\in H_j\}=Z(H_j)$,
whose Lie algebra is $\Lh_j$. So we can form the maps
$\varphi^\grp_j:G/Z(H_j)\times H_j\rightarrow G_j$ and
$\varphi^\alg_j:G/H_j\times \Lh_j\rightarrow \Lg_j$ by
$\varphi^\grp_j([g],h)=\sigma_g(h)$ and
$\varphi^\alg_j([g],\eta)=\Ad_g(\eta)$, respectively. The maps
$\varphi^\grp_j$ and $\varphi^\alg_j$ may not be surjective in
general. But since $G'_j\subset\IM(\varphi^\grp_j)\subset G_j$ and
$\Lg'_j\subset\IM(\varphi^\alg_j)\subset\Lg_j$, the sets
$G_j\backslash\IM(\varphi^\grp_j)$ and
$\Lg_j\backslash\IM(\varphi^\alg_j)$ have measure zero. Choose a
Hermitian product on the complexification $\Lg_\CC$ of $\Lg$ such
that $(\Lh_j)_\CC$ and the associated root spaces are mutually
orthogonal. It induces a left invariant Riemannian structure on
$G$, then induces a $G$-invariant measure $dg_j$ on $G_j$ and a
Haar measure $dh_j$ on $H_j$. Note that the measure $dg_j$ is the
restriction of a Haar measure $dg$ on $G$ for each $j$. Similarly,
it induces a $G$-invariant measure $dX_j$ on $\Lg_j$ (which is the
restriction of a Lebesgue measure $dX$ on $\Lg$ for each $j$) and
a Lebesgue measure $dY_j$ on $\Lh_j$. Let $d\mu'_j, d\mu_j$ be
$G$-invariant measures on $G/Z(H_j)$ and $G/H_j$, which are
induced by Riemannian structures on $G/Z(H_j)$ and $G/H_j$ such
that the identifications $\Lh_j^\bot\cong T_{[e]}(G/Z(H_j))\cong
T_{[e]}(G/H_j)$ are isometric. For each
$\alpha\in\Delta_j=\Delta(\Lg_\CC,(\Lh_j)_\CC)$, Let
$\vartheta_{\alpha}$ be the restriction of the adjoint
representation of $H_j$ on the root space $\Lg_\alpha$, which
satisfies $\vartheta_{\alpha}(e^\eta)=e^{\alpha(\eta)}$ for
$\eta\in\Lh_j$. Then we have

\begin{theorem}\label{T:reductive-group}
Let the objects be as above. Then for each $1\leq j\leq m$,\\
(1) $(G,\sigma,G_j,dg_j,H_j,dh_j)$ is a generalized random matrix
ensemble. Its generalized joint density function
$\mathcal{P}_j(h)=J^\grp_j(h)$ is given by
{\small\begin{equation}\label{E:reductive-group}
J^\grp_j(h)=\prod_{\alpha\in\Delta_j}|1-\vartheta_{\alpha}(h^{-1})|.
\end{equation}}
(2)$(G,\Ad,\Lg_j,dX_j,\Lh_j,dY_j)$ is a generalized random matrix
ensemble. Its generalized joint density function
$\mathcal{P}_j(\eta)=J^\alg_j(\eta)$ is given by
{\small\begin{equation}\label{E:reductive-algebra}
J^\alg_j(\eta)=\prod_{\alpha\in\Delta_j}|\alpha(\eta)|.
\end{equation}}
\end{theorem}

\begin{proof}
First we prove (1). We let $(G_j)_\z=G_j\backslash G'_j$,
$(H_j)_\z=H_j\backslash H'_j$. Then by the discussions above, the
condition (a) holds automatically. For $h\in H'_j$, since $h$ is
regular, the Lie algebra of $G_h=\{g\in G:ghg^{-1}=h\}$ is
$\Lh_j$, so the dimension condition (c) holds. For $h\in H_j$ and
$\xi\in\Lh_j^\bot$, under the identification of $T_hG$ with
$\Lg=T_eG$ by left multiplication, it is easy to show that
{\small$$ \Psi_h(\xi)=\frac{d}{dt}\Big|_{t=0}\sigma_{\exp
t\xi}(h)=(\Ad(h^{-1})-I)\xi.
$$}
Let
$\Lg_\CC=(\Lh_j)_\CC\oplus\bigoplus_{\alpha\in\Delta_j}\Lg_\alpha$
be the root space decomposition of $\Lg_\CC$, then
$(\Lh_j^\bot)_\CC=\bigoplus_{\alpha\in\Delta_j}\Lg_\alpha$. Let
$\Delta_j^1=\{\alpha\in\Delta_j:\Lh_j^\bot\cap\Lg_\alpha\neq0\}$,
and let $\Delta_j^2=\Delta_j\backslash\Delta_j^1$. For each
$\alpha\in\Delta_j$, choose a $\xi_\alpha\in\Lg_\alpha$ such that
$\xi_\alpha\in\Lh_j^\bot$ for $\alpha\in\Delta_j^1$. Then for
$\alpha\in\Delta_j^1$, {\small\begin{equation}\label{1}
\Psi_h(\xi_\alpha)=(\vartheta_{\alpha}(h^{-1})-1)\xi_\alpha.
\end{equation}}
Now let $\alpha\in\Delta_j^2$. For each $h\in H_j$, since
$\Ad(h)\xi_\alpha=\vartheta_{\alpha}(h)\xi_\alpha$, we have
$\Ad(h)\overline{\xi_\alpha}=
\overline{\vartheta_{\alpha}(h)}\;\overline{\xi_\alpha}$, where
$\overline{\xi_\alpha}$ is the conjugation $\xi_\alpha$ of with
respect to $\Lh_j$. This means that $\overline{\xi_\alpha}$
belongs to some root space $\Lg_{\alpha'}$. Denote
$\alpha'=\tau(\alpha)$, then $\tau$ is a permutation of
$\Delta_j^2$ without fixed point, and $\tau^2=1$. So $\Delta_j^2$
has a partition
$\Delta_j^2=\widetilde{\Delta}_j^2\sqcup\tau(\widetilde{\Delta}_j^2)$.
Modifying the Hermitian product on $\Lg_\CC$ if necessary, we may
assume $|\xi_\alpha|=|\overline{\xi_\alpha}|$. Then it is easy to
show that
$\{\xi_\alpha:\alpha\in\Delta_j^1\}\cup\{\xi_\alpha+\overline{\xi_\alpha},
i(\xi_\alpha-\overline{\xi_\alpha}):\alpha\in\widetilde{\Delta}_j^2\}$
is an orthogonal basis of $\Lh_j^\bot$. Now for
$\alpha\in\widetilde{\Delta}_j^2$, we have
{\small\begin{equation}\label{2}
\Psi_h(\xi_\alpha+\overline{\xi_\alpha})
=(\vartheta_{\alpha}(h^{-1})-1)\xi_\alpha+
\overline{(\vartheta_{\alpha}(h^{-1})-1)}\;\overline{\xi_\alpha},
\end{equation}
\begin{equation}\label{3}
\Psi_h(i(\xi_\alpha-\overline{\xi_\alpha}))
=i(\vartheta_{\alpha}(h^{-1})-1)\xi_\alpha-
i\overline{(\vartheta_{\alpha}(h^{-1})-1)}\;\overline{\xi_\alpha}.
\end{equation}}
If $h\in H'_j$, then $\vartheta_{\alpha}(h^{-1})-1\neq0,
\forall\alpha\in\Delta_j$. This means that
$\Psi_h:\Lh_j^\bot\rightarrow T_hO_h$ is an isomorphism, and
$T_hG'_j=\Lh_j\oplus T_hO_h$ orthogonally. So the conditions (b)
and (d) hold. Combining \eqref{1}, \eqref{2}, and \eqref{3}, we
get
{\small$$J^\grp_j(h)=|\det\Psi_h|=\prod_{\alpha\in\Delta_j}|1-\vartheta_{\alpha}(h^{-1})|.$$}
This proves (1). The proof of (2) is similar, which was omitted
here.
\end{proof}

\begin{corollary}\label{C:reductive-groupInt}
Let the objects be as above. Then for each $1\leq j\leq m$, we
have {\small\begin{equation}\label{E:reductive-groupInt}
\int_{G_j} f(g)
dg_j=\frac{1}{|W_j|}\int_{H_j}\left(\int_{G/H_j}f(\sigma_g(h))d\mu_j([g])\right)J^\grp_j(h)
dh_j,
\end{equation}
\begin{equation}\label{E:reductive-algebraInt}
\int_{\Lg_j} f(\xi)
dX_j(\xi)=\frac{1}{|W_j|}\int_{\Lh_j}\left(\int_{G/H_j}f(\Ad_g(\eta))d\mu_j([g])\right)J^\alg_j(\eta)
dY_j(\eta),
\end{equation}}
where $W_j$ is the analytic Weyl group $W_j=N_G(\Lh_j)/H_j$
associated with $H_j$.
\end{corollary}

\begin{proof}
The proof of \eqref{E:reductive-algebraInt} is essentially same to
the proof of \eqref{E:complex-algebraInt} in Corollary
\ref{C:complex-groupInt}. Now we proof formula
\eqref{E:reductive-groupInt}. Similar to the proof of
\eqref{E:complex-groupInt} in Corollary \ref{C:complex-groupInt},
we can get an integration formula
\begin{equation}\label{E:reductive-groupInt2}
\small{\int_{G_j} f(g)
dg_j=\\
\frac{1}{|N_G(H_j)/Z_G(H_j)|}\int_{H_j}\left(\int_{G/Z(H_j)}f(\sigma_g(h))d\mu'_j([g])\right)J^\grp_j(h)
dh_j.}
\end{equation}
But $Z_G(H_j)=Z(H_j)$, and it is easily to prove
$N_G(H_j)=N_G(\Lh_j)$. So $|N_G(H_j)/Z_G(H_j)|$
$=|N_G(\Lh_j)/Z(H_j)|=|N_G(\Lh_j)/H_j|\cdot|H_j/Z(H_j)|=|W_j|\cdot|H_j/Z(H_j)|$.
Hence to prove \eqref{E:reductive-groupInt}, by
\eqref{E:reductive-groupInt2}, it is sufficient to show that
{\small\begin{equation}\label{E:reductive-groupInt3}
\int_{G/Z(H_j)}f(\sigma_g(h))d\mu'_j([g])=|H_j/Z(H_j)|\int_{G/H_j}f(\sigma_g(h))d\mu_j([g]).
\end{equation}}
But the natural projection $\psi:G/Z(H_j)\rightarrow G/H_j$ is a
$|H_j/Z(H_j)|$-sheeted covering map, and $\psi^*(d\mu_j)=d\mu'_j$.
Hence \eqref{E:reductive-groupInt3} follows directly from
Proposition 3.1 in \cite{AWY}. This complete the proof of the
corollary.
\end{proof}

\begin{corollary}\label{C:reductive-groupInt4}
Let the objects be as above. Then we have
{\small\begin{equation}\label{E:reductive-groupInt4} \int_G f(g)
dg=\sum_{j=1}^m\frac{1}{|W_j|}\int_{H_j}\left(\int_{G/H_j}f(\sigma_g(h))d\mu_j([g])\right)J^\grp_j(h)
dh_j,
\end{equation}
\begin{equation}\label{E:reductive-algebraInt4}
\int_\Lg f(\xi)
dX(\xi)=\sum_{j=1}^m\frac{1}{|W_j|}\int_{\Lh_j}\left(\int_{G/H_j}f(\Ad_g(\eta))d\mu_j([g])\right)J^\alg_j(\eta)
dY_j(\eta).
\end{equation}}
\end{corollary}

\begin{proof}
Since $G'=\bigsqcup_{j=1}^mG'_j$, $\Lg'=\bigsqcup_{j=1}^m\Lg'_j$,
and the sets of singular elements $G_s=G\backslash G'$ and
$\Lg_s=\Lg\backslash\Lg'$ have measure zero, and also notice that
$H_j\backslash H'_j$ and $\Lh_j\backslash\Lh'_j$ have measure zero
in the corresponding spaces, the proof follows directly from
Corollary \ref{C:reductive-groupInt}.
\end{proof}

\begin{remark}
Formula \eqref{E:reductive-groupInt4} is just the Harish-Chandra's
integration formula for real reductive groups (see \cite{Kn},
Theorem 8.64), and formula \eqref{E:reductive-algebraInt4} is its
linear version. Here we recover them from the viewpoint of
generalized random matrices.
\end{remark}

\begin{example}\label{Ex:SL(2,R)}
Let $G=SL(2,\RR)$. Its Lie algebra $\Lg=\Lsl(2,\RR)=\left\{{\SMALL
\begin{pmatrix}
x & y\\
z & -x
\end{pmatrix}}\right\},$
where $x,y,z\in\RR$. $\theta {\SMALL\begin{pmatrix}
x & y\\
z & -x
\end{pmatrix}}
=-{\SMALL\begin{pmatrix}
x & y\\
z & -x
\end{pmatrix}}^t
={\SMALL\begin{pmatrix}
-x & -z\\
-y & x
\end{pmatrix}}
$ is a Cartan involution of $\Lsl(2,\RR)$. There are exactly $2$
mutually nonconjugate $\theta$ stable Cartan subalgebras
$\Lh_1=\left\{{\SMALL
\begin{pmatrix}
x & 0\\
0 & -x
\end{pmatrix}}\right\}, \Lh_2=\left\{{\SMALL
\begin{pmatrix}
0 & y\\
-y & 0
\end{pmatrix}}\right\}.$ The corresponding Cartan subgroups are
$H_1=\left\{{\SMALL
\begin{pmatrix}
a & 0\\
0 & a^{-1}
\end{pmatrix}}:a\in\RR,a\neq0\right\},H_2=SO(2)$ (see \cite{Kn}, page
487). Note that $H_2$ is connected, but $H_1$ has two connected
components. It is easy to show that
$$\Lg'=\Lg_r=\{\xi\in\Lsl(2,\RR):\det\xi\neq0\},$$
$$\Lg_1=\{\xi\in\Lsl(2,\RR):\det\xi\leq0\},$$
$$\Lg_2=\{\xi\in\Lsl(2,\RR):\det\xi\geq0\}.$$
Similarly,
$$G'=G_r=\{g\in SL(2,\RR):|\tr\; g|\neq2\},$$
$$G_1=\{g\in SL(2,\RR):|\tr\; g|\geq2\},$$
$$G_2=\{g\in SL(2,\RR):|\tr\;g|\leq2\}.$$
The corresponding root systems are $\Delta_1=\{\pm\alpha_1\}$ with
$\pm\alpha_1
{\SMALL
\begin{pmatrix}
x & 0\\
0 & -x
\end{pmatrix}}
=\pm 2x$, and $\Delta_2=\{\pm\alpha_2\}$ with $\pm\alpha_2{\SMALL
\begin{pmatrix}
0 & y\\
-y & 0
\end{pmatrix}}
=\pm 2iy$. By Theorem \ref{T:reductive-group}, the density
function $\mathcal{P}_1=J^\alg_1$ for the pseudo-algebra ensemble
$(SL(2,\RR),\Ad,\Lg_1,dX_1,\Lh_1,dY_1)$ is
\begin{equation}
J^\alg_1{\SMALL\begin{pmatrix}
x & 0\\
0 & -x
\end{pmatrix}}
=4x^2,
\end{equation}
and the density function $\mathcal{P}_2=J^\alg_2$ for the
pseudo-algebra ensemble $(SL(2,\RR),\Ad,\Lg_2,$ $dX_2,\Lh_2,dY_2)$
is
\begin{equation}
J^\alg_2{\SMALL\begin{pmatrix}
0 & y\\
-y & 0
\end{pmatrix}}
=4y^2.
\end{equation}
For the group ensembles, it is easy to show that
$\vartheta_{\pm\alpha_1}{\SMALL
\begin{pmatrix}
a & 0\\
0 & a^{-1}
\end{pmatrix}}
=a^{\pm2}$, and \\
$\vartheta_{\pm\alpha_2}{\SMALL
\begin{pmatrix}
\cos y & \sin y\\
-\sin y & \cos y
\end{pmatrix}}
=e^{\pm2iy}$. So by Theorem \ref{T:reductive-group}, the density
function $\mathcal{P}_1=J^\grp_1$ for the pseudo-group ensemble
$(SL(2,\RR),\sigma,G_1,dg_1,H_1,dh_1)$ is
\begin{equation}
J^\grp_1{\SMALL\begin{pmatrix}
a & 0\\
0 & a^{-1}
\end{pmatrix}}
=\left(a-a^{-1}\right)^2,
\end{equation}
and the density function $\mathcal{P}_2=J^\grp_2$ for the
pseudo-group ensemble $(SL(2,\RR),\sigma,G_2,$ $dg_2,H_2,dh_2)$ is
\begin{equation}
J^\grp_2 {\SMALL\begin{pmatrix}
\cos y & \sin y\\
-\sin y & \cos y
\end{pmatrix}}
=4\sin^2y.
\end{equation}\qed
\end{example}

\begin{example}\label{Ex:GL(n,R)}
Let $G=GL(n,\RR)$, which is a reductive group. Then
$\theta(\xi)=-\xi^t$ is a Cartan involution of $\Lg=\gl(n,\RR)$.
$\Lg$ has exactly $m=\left[\frac{n}{2}\right]+1$ mutually
nonconjugate $\theta$ stable Cartan subalgebras, which can be
write explicitly as
$$
\Lh_j=\left\{\eta=D_j(x_1\cdots x_n):x_k\in\RR\right\},
$$
$j=0,\cdots,\left[\frac{n}{2}\right]$, where we denote
$$
D_j(x_1\cdots x_n)= {\SMALL\begin{pmatrix}
\diag(x_1\cdots x_j) & \diag(x_{j+1}\cdots x_{2j}) & 0\\
-\diag(x_{j+1}\cdots x_{2j}) & \diag(x_1\cdots x_j) & 0\\
0 & 0 & \diag(x_{2j+1}\cdots x_n)
\end{pmatrix}}
$$
(see \cite{Wa}, page 95). Using the explicit form of the Cartan
subalgebra $\Lh_j$, one can easily prove that an $n$-by-$n$ real
matrix commutes with all elements in $\Lh_j$ if and only if it is
of the form $D_j(a_1\cdots a_n)$, whose determinant is
$\prod_{r=1}^j(a_r^2+a_{j+r}^2)\prod_{r=2j+1}^na_r$. So by
definition, the Cartan subgroup
{\small$$
H_j=\Big\{h=D_j(a_1\cdots a_n) :a_k\in\RR,
\prod_{r=1}^j(a_r^2+a_{j+r}^2)\prod_{r=2j+1}^na_r\neq0 \Big\}.
$$}
It is easily seen that $H_j$ has $2^{n-2j}$ components. The
integration manifolds $\Lg_j=\overline{\bigcup_{g\in
G}\Ad_g(\Lh_j)}$, $G_j=\overline{\bigcup_{g\in G}gH_jg^{-1}}$.
More precisely, one can prove that
$$\Lg'_j=\{\xi\in\Lg':\xi\;\text{has exactly}\; n-2j \;\text{real eigenvalues}\},$$
$$G'_j=\{g\in G':g\;\text{has exactly}\; n-2j \;\text{real eigenvalues}\}.$$
So
\begin{align*}
\Lg_j=\overline{\Lg'_j}=\{\xi\in&\Lg:\;\text{for some suitable
permutation}\; \lambda_1,\cdots,\lambda_n \\\;\text{of the
eigenvalues of
}&\;\xi,\;\lambda_1=\overline{\lambda_{j+1}},\cdots,\lambda_j=\overline{\lambda_{2j}}\;;\;
\lambda_{2j+1},\cdots,\lambda_n\;\text{are real}\},\\
G_j=\overline{G'_j}=\{g\in& G:\;\text{for some suitable
permutation}\; \lambda_1,\cdots,\lambda_n \\\;\text{of the
eigenvalues of
}&\;g,\;\lambda_1=\overline{\lambda_{j+1}},\cdots,\lambda_j=\overline{\lambda_{2j}}\;;\;
\lambda_{2j+1},\cdots,\lambda_n\;\text{are real}\}.
\end{align*}
The root system associated with the Cartan subalgebra
$\Lh_0=\{\diag(x_1,\cdots,x_n)\}$ is
$\Delta_0=\{\pm(\ee_r-\ee_s):1\leq r<s\leq n\},$ where
$\ee_r\in\Lh_0^*$ is defined by $\ee_r(\diag(x_1,\cdots,x_n))$
$=x_r$. Denote the matrix $L={\SMALL\begin{pmatrix}
I_j & I_j & 0\\
iI_j & -iI_j & 0\\
0 & 0 & I_{n-2j}
\end{pmatrix}},$
Then we have $L^{-1}D_j(x_1\cdots x_n)L$ $=\diag(y_1,\cdots,y_n),$
where {\small$$ y_r=\begin{cases}
x_r+ix_{j+r}, &1\leq r\leq j;\\
x_{r-j}-ix_{r}, &j+1\leq r\leq 2j;\\
x_r, &2j+1\leq r\leq n.
\end{cases}$$}
By Theorem \ref{T:reductive-group}, the density function
$\mathcal{P}_j(\eta)=J^\alg_j(\eta)$ for the pseudo-algebra
ensemble $(GL(n,\RR),\Ad,\Lg_j,dX_j,\Lh_j,dY_j)$ is
{\small\begin{align}\label{F:gl} J^\alg_j(\eta)
=&\prod_{\alpha\in\Delta_0}\big|\alpha\big(\diag(y_1,\cdots,y_n)\big)\big|\notag\\
=&\prod_{1\leq r<s\leq n}|y_r-y_s|^2\notag\\
=&\prod_{1\leq r<s\leq j}\big|[(x_r+ix_{j+r})-(x_s+ix_{j+s})]
[(x_r-ix_{j+r})-(x_s-ix_{j+s})]\big|^2\notag\\
&\prod_{2j+1\leq r<s\leq n}|x_r-x_s|^2
\prod_{1\leq r,s\leq j}\big|(x_r+ix_{j+r})-(x_s-ix_{j+s})\big|^2\\
&\prod_{1\leq r\leq j,2j+1\leq s\leq
n}\big|[(x_r+ix_{j+r})-x_s][(x_r-ix_{j+r})-x_s]\big|^2\notag\\
=&\;4^j\prod_{r=1}^jx_{j+r}^2\prod_{2j+1\leq r<s\leq
n}|x_r-x_s|^2\prod_{1\leq r\leq j,2j+1\leq s\leq
n}\big((x_r-x_s)^2+x_{j+r}^2\big)^2\notag\\
&\prod_{1\leq r<s\leq
j}\big((x_r-x_s)^2+(x_{j+r}-x_{j+s})^2\big)^2\big((x_r-x_s)^2+(x_{j+r}+x_{j+s})^2\big)^2.\notag
\end{align}}

\vskip 0.3cm Now we come to the groups ensembles associated with
$G=GL(n,\RR)$. A direct computation shows that the root spaces
$\Lg_\alpha (\alpha\in\Delta_j)$ associated with
$(\Lg_\CC,(\Lh_j)_\CC)$ are of the form
$$\{\Lg_\alpha:\alpha\in\Delta_j\}=\{\CC
(LE_{rs}L^{-1}):r\neq s\},$$ where $E_{rs}$ denotes the $n$-by-$n$
matrix with $1$ at the $(r,s)$ position and $0$ elsewhere. We
denote the root $\alpha\in\Delta_j$ corresponding to $E_{rs}$ by
$\alpha_{rs}$. One can also easily computes that for each $h\in
H_j$ and $\xi_{rs}\in\Lg_{\alpha_{rs}}$, $h\xi_{rs}
h^{-1}=\frac{l_r}{l_s}\xi_{rs}$, where
{\small $$l_r=\begin{cases}
h_r+ih_{j+r}, &1\leq r\leq j;\\
h_{r-j}-ih_{r}, &j+1\leq r\leq 2j;\\
h_r, &2j+1\leq r\leq n.
\end{cases}$$}
So $\vartheta_{\alpha_{rs}}(h)=\frac{l_r}{l_s}$, and then by
Theorem \ref{T:reductive-group},
{\small\begin{align*} J^\grp_j(h)
=&\prod_{1\leq r,s\leq n;r\neq s}\Big|1-\vartheta_{\alpha_{rs}}(h^{-1})\Big|\\
=&\prod_{1\leq r<s\leq
n}\Big|1-\frac{l_r}{l_s}\Big|\Big|1-\frac{l_s}{l_r}\Big|\\
=&\prod_{1\leq r<s\leq n}\frac{|l_r-l_s|^2}{|l_rl_s|}.
\end{align*}}
Note that the expression $\prod_{1\leq r<s\leq n}|l_r-l_s|^2$ has
been computed in Formula \eqref{F:gl}, if we replace $y_r$ by
$l_r$. On the other hand, {\small$$\prod_{1\leq r<s\leq
n}|l_rl_s|=\prod_{r=1}^{n}|l_r|^{n-1}=\prod_{r=1}^j
(h_r^2+h_{j+r}^2)^{n-1}\prod_{r=2j+1}^n|h_r|^{n-1}.$$} Combining
these two results, we get {\small\begin{align}\label{F:GL2}
J^\grp_j(h)=&\;4^j\prod_{r=1}^jh_{j+r}^2(h_r^2+h_{j+r}^2)^{-(n-1)}
\prod_{r=2j+1}^n|h_r|^{-(n-1)}\prod_{2j+1\leq r<s\leq
n}|h_r-h_s|^2\notag\\
&\prod_{1\leq r<s\leq
j}\big((h_r-h_s)^2+(h_{j+r}-h_{j+s})^2\big)^2
\big((h_r-h_s)^2+(h_{j+r}+h_{j+s})^2\big)^2\\
&\prod_{1\leq r\leq j,2j+1\leq s\leq
n}\big((h_r-h_s)^2+h_{j+r}^2\big)^2.\notag
\end{align}}\qed
\end{example}

\end{document}